\documentclass[sigconf]{acmart}

\usepackage{amsmath,amsfonts}
\usepackage{algorithmic}
\usepackage{graphicx}
\usepackage{textcomp}
\usepackage{xcolor}
\usepackage{verbatim}
\usepackage{tabularx}
\usepackage{xspace}
\usepackage{multirow}
\usepackage{subcaption}
\usepackage{booktabs}
\usepackage{soul}
\usepackage{url}
\usepackage{hyperref}
\usepackage{enumitem}
\usepackage{colortbl}

\definecolor{l-gray}{gray}{0.85}

\newcommand{\updatetext}[1]{\textcolor{black}{#1}}

\copyrightyear{2022}
\acmYear{2022}
\setcopyright{rightsretained}
\acmConference[ASE '22]{37th IEEE/ACM International Conference on
Automated Software Engineering}{October 10--14, 2022}{Rochester, MI,
USA}
\acmBooktitle{37th IEEE/ACM International Conference on Automated
Software Engineering (ASE '22), October 10--14, 2022, Rochester, MI,
USA}
\acmDOI{10.1145/3551349.3556925}
\acmISBN{978-1-4503-9475-8/22/10}

\begin{document}

\title{Data Augmentation for Improving Emotion Recognition in Software Engineering Communication}

\author{Mia Mohammad Imran}
\affiliation{%
  \institution{Virginia Commonwealth University}
  \city{Richmond, Virginia}
  \country{USA}
}
\email{imranm3@vcu.edu}

\author{Yashasvi Jain}
\affiliation{%
  \institution{Drexel University}
  \city{Philadelphia, Pennsylvania}
  \country{USA}
}
\email{yj395@drexel.edu}

\author{Preetha Chatterjee}
\affiliation{%
  \institution{Drexel University}
  \city{Philadelphia, Pennsylvania}
  \country{USA}
}
\email{pc697@drexel.edu}

\author{Kostadin Damevski}
\affiliation{%
  \institution{Virginia Commonwealth University}
  \city{Richmond, Virginia}
  \country{USA}
}
\email{kdamevski@vcu.edu}

\begin{abstract}
Emotions (e.g., Joy, Anger) are prevalent in daily software engineering (SE) activities, and are known to be significant indicators of work productivity (e.g., bug fixing efficiency). Recent studies have shown that directly applying general purpose emotion classification tools to SE corpora is not effective. Even within the SE domain, tool performance degrades significantly when trained on one communication channel and evaluated on another (e.g, StackOverflow vs. GitHub comments). 
Retraining a tool with channel-specific data takes significant effort since manually annotating a large dataset of ground truth data is expensive.

In this paper, we address this data scarcity problem by automatically creating new training data using a data augmentation technique. Based on an analysis of the types of errors made by popular SE-specific emotion recognition tools, we specifically target our data augmentation strategy in order to improve the performance of emotion recognition.   
Our results  show an average improvement of 9.3\% in micro F1-Score for three existing emotion classification tools (ESEM-E, EMTk, SEntiMoji) when trained with our best augmentation strategy. 
\end{abstract}

\keywords{}

\maketitle

\section{Introduction}

% \textbf{Overall big research problem}
Emotions can strongly impact activities that are collaborative in nature and require creativity and problem-solving skills, such as software development~\cite{Amabile2005AffectAC}. 
Recent research has shown that positive emotions (e.g., Joy) are associated with increased productivity and job satisfaction in software engineering teams \cite{Graziotin13, Girardi20, Muller15, 8666786, Forsgren21}. On the other hand, negative emotions (e.g., Frustration) can cause developers to lose motivation and exhibit lower participation, ultimately leading to team attrition \cite{GRAZIOTIN201832}. 
Negative emotions, which are known to impact cognitive processes, could also serve as an obstacle to learning a new programming language, code comprehension, etc.~\cite{8802324}.
Thus, for quite a while, software engineering researchers have studied developer emotions and how they impact software development activities~\cite{Tourani, ortu2015bullies, sinha2016analyzing}.
The goals of the research have been to empirically understand the causes and the impact of different emotional states on the productivity of an individual
developer or a team~\cite{Graziotin15, Muller15, Graziotin17, GRAZIOTIN201832} and to design techniques that automatically detect developer emotion and provide recommendations to developers~\cite{Sarker19, Kuutila2020ChatAI, Deva, Fucci21, Ebert2021AnES,8802324, Chatterjee21}. 
Success in achieving these goals is predicated on the ability to detect emotions with high accuracy. 

%%% https://www.win.tue.nl/~aserebre/TOSEM2021.pdf
%1. Tool performance varies on different data.
%2. Retraining a tool with software-related data requires substantial effort
%%% implicitly, both are about data scarcity
While several approaches that target emotion detection in software developers' written text have been proposed, they have not been evaluated extensively due to the limited availability of manually annotated ground truth data~\cite{opinionlitreview}. Manual annotation of emotions in text is time consuming and requires additional effort to ensure annotator subjectivity is minimized~\cite{Imtiaz18, opinionlitreview}. At the same time, there is ample evidence that emotion classifiers trained on general purpose data (i.e., not from software engineering) perform poorly in the software engineering domain, likely due to the specific vocabulary and characteristics of software engineering communication (e.g., occasional presence of short snippets of code)~\cite{novielli2021assessment,Senti4SD}. In fact, emotion classification performance appears to be optimal when trained and evaluated on each specific software engineering channel separately, e.g., GitHub issue comments, StackOverflow comments~\cite{8595220}. 

In this paper, our aim is to understand the current limitations and improve emotion classification in software engineering written communication. 
More specifically, we address the following two research questions:

\smallskip 
\noindent
{\bf RQ1:} {\em How effective are existing emotion classifiers in detecting emotions in GitHub comments?
What types of emotions are most likely to be misclassified?}

To answer this RQ, we first create a new dataset for emotion classification based on GitHub issue and pull request discussions. We annotate the dataset based on the six emotion categories first introduced by Shaver\cite{Shaver}, while also going beyond this categorization to a finer grained division using secondary and tertiary emotions. Using our dataset, we evaluate three of the most commonly used tools for software engineering emotion classification (ESEM-E~\cite{esem-e}, EMTk~\cite{emtk}, SEntiMoji~\cite{sentimoji}), showing that their accuracy is further reduced compared to the original datasets that the tools were built and evaluated for. We perform an error analysis focused on the instances that all of the three tools got wrong, in order to understand the limitations of the current generation of approaches. Our results indicate that a large number of the errors (the majority) are due to a simple inability of the tools to recognize clear lexical cues that are present in the text, and not due to more hard-to-discern causes, such as the emotion being implicitly expressed.

\smallskip 
\noindent
{\bf RQ2:} {\em Can automatic data augmentation techniques be used to improve the effectiveness of existing emotion classifiers?}

To answer this RQ, we explore three different strategies for data augmentation. Each of the strategies significantly increases the training set size, generating 10x more training data than what we had at the outset. The strategies contrast between unconstrained augmentation, where we modify the original text in random places and without additional checks, and augmentation with a number of constraints that ensure it does not perturb the original emotion in the text.
Our experimental results show that the best strategy focuses on preserving the emotional polarity of the text, which is positive for emotions like Joy and Love and negative for emotions like Anger. Focusing the augmentation directly on individual emotions is not as profitable, as we lack sufficiently large software engineering-specific lexicons that can be used to generate a large and diverse number of augmented instances.

\vspace{0.2cm}

% \updatetext{ {\bf Contributions:} The key contribution of this paper are: 
% \begin{itemize}
%     \item operationalization (extension) of an emotion taxonomy for textual SE  communication artifacts;
%     \item an annotated dataset of Github pull requests and issue comments to evaluate three existing emotion detection tools for SE artifacts;
%     \item quantitative and qualitative evaluation along with the error analysis to understand the limitations of existing emotion detection tools in SE;
%     \item three data augmentation approaches to address the data scarcity issue of automatic emotion detection in SE-related text; 
%     \item evaluation of these data augmentation approaches with the aforementioned tools.
% \end{itemize}
% }

{\bf Contributions:} Improved emotion recognition \updatetext{taxonomy} in developer written communication can impact empirical research as well as motivate new tools that improve emotion awareness in software engineering projects. To the best of our knowledge, we are the first to explore data augmentation strategies in order to deal with the scarcity of high-quality annotated data and improve emotion recognition in software engineering-related text. A similar approach to ours could potentially also be leveraged to improve related tasks in software engineering, such as automatic generation of training data for sentiment analysis. 

We publish the annotation instructions, annotated dataset, and source code to facilitate the replication of our study at:  \url{https://anonymous.4open.science/r/SE-Emotion-Study-0141/}

% The major contributions of this paper are:
% \begin{itemize}[leftmargin=*]
%     \item A dataset of 2000 Github issue report and pull request comments, manually annotated using modified Shaver's emotion model~\cite{shaver}.
%     \item A set of data augmentation techniques to ...
%     \item We evaluated our techniques on Github  comments from four different programming communities. Our evaluation results show that ...
%     \item 
% \end{itemize}

\section{Background}

Over the years, emotions have been conceptualized in different ways by software engineering researchers. We begin by providing background on measuring emotions and the existing annotated datasets. We then introduce data augmentation, a technique for dealing with data scarcity in machine learning by increasing the training set size.

\subsection{Emotions in Software Artifacts} 
Leveraging research in psychology, over the years researchers have used several categorizations of emotions in written text. For instance, Ekman et al.~\cite{ekman} categorized emotions into six basic categories:  Anger, Disgust, Fear, Joy, Sadness, and Surprise. 
On the other hand, Shaver et al.~\cite{Shaver} identified six basic emotion categories: Love, Joy, Anger, Sadness, Surprise, and Fear. Shaver et al. expanded the basic set of emotions to secondary and tertiary levels in a tree-like structure. Shaver's emotional structure was refined in the work of Parrott et al.~\cite{parrott2001emotions}. Plutchik~\cite{PLUTCHIK19803} proposed a wheel-like structure of three layers of emotions with eight basic categories: Anger, Disgust, Fear, Joy, Sadness, Surprise, Trust, and Anticipation.
Studies conducted by Cowen et al. identified 27 distinct categories based on 2185 short videos~\cite{CowenE7900}, 28 categories using facial expressions~\cite{cowen2020face}, and 24 using human vocalization~\cite{cowen2019mapping}. Based on Cowen et al.'s studies, Demszky et al. devised 27 categories for text-based emotion recognition~\cite{goemotions}. They also provided a mapping between these 27 categories and Ekman et al.'s six basic categories~\cite{ekman}. 
Although it is widely accepted that emotions are comprised of basic categories that are combined to form more complex emotions (e.g, Frustration), there is no consensus on the complete list of categories that accommodate the wide range of emotions observed in the
software engineering domain~\cite{sanei2021impacts}.% \updatetext{as well as other domains~\cite{PLUTCHIK19803}}. 
% In this paper, we leverage Shaver et al. categories~\cite{shaver} as they are the most commonly supported by the current generation of emotion recognition tools in software engineering.

\textbf{\textit{Available Datasets.}} To conduct studies of software developer emotions in various communication channels and software artifacts, researchers have created a few manually annotated datasets (i.e., gold sets) that leverage the categories described above. For instance, Ortu et al.~\cite{ortu2016} annotated a JIRA dataset that contained 5992 issue samples in three groups: group 1 contained 392 issue comments labeled with emotions Love, Joy, Surprise, Anger, Fear and Sadness; group 2 contained 1600 issue comments labeled with emotions Love, Joy, and
Sadness; and group 3 contained 4000 issue sentences labeled with emotions Love, Joy, Anger, and Sadness.
Novielli et al.~\cite{novielli2018gold} annotated a gold set from 4800 StackOverflow questions, answers, and comments. They labeled the sentences with Shaver et al~\cite{Shaver}'s six basic categories.
% Islam et al.~\cite{Deva} annotated a Jira dataset based on VAD model~\cite{russell1977evidence} which is different from Ortu et al.~\cite{ortu2016}'s Jira dataset. This dataset contained 1755 issue comments.
Venigalla et al.~\cite{venigalla-emotion} analyzed 10996 commit messages related to software documentation update from 998 GitHub projected and mapped them into Plutchik's eight emotion categories~\cite{PLUTCHIK19803}.

\begin{figure}
\centering
\includegraphics[width=0.9\linewidth]{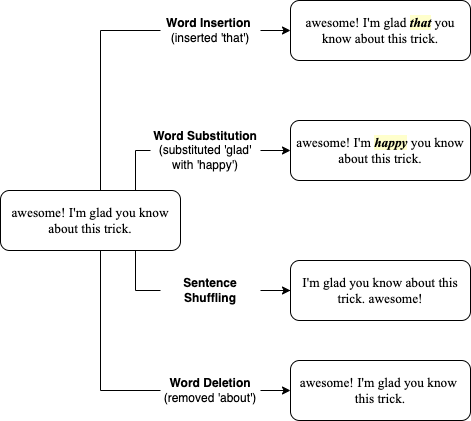}
\caption{Example of data augmentation using four  operators.}
\vspace{-0.4cm}
\label{fig:data_augmentation}
\end{figure}

\subsection{Data Augmentation}

Data augmentation (DA)~\cite{feng2021survey, li2022data} is a technique for increasing training data diversity by targeted modification of the existing data. Insufficient quantity of high-quality training data is a common problem in machine learning, especially with the emergence of more complex models, e.g., deep learning. The concept of DA originates in image processing, where researchers observed that, e.g., rotating an image by 90 degrees produces a new training instance for image classification tasks that increases the robustness of the model. Applying DA to written text has recently gained popularity despite the fact that this is usually a more difficult problem due to the complex relationship among written words. More specifically, DA has to maintain {\em label invariance}, that is, the augmented instance needs to have the same label as the original instance. 

Data augmentation works by applying augmentation operators, one or more times to each training set instance. Popular and simple DA operators for text are word insertion, word substitution, sentence shuffling, word deletion, etc. Figure~\ref{fig:data_augmentation} shows an example data augmentation of an utterance from a GitHub issue comment using these operators. However, research shows that operators that target the specific classification task (i.e., emotion detection) are significantly more effective than generic ones~\cite{kovatchev2021vectors}. Researchers also often use backtranslation (i.e., automatically translating to another language and back to English) and large language models (e.g., BERT) as DA operators, which are more likely to significantly change the original text, but with more risk towards perturbing the classification label.

\section{Methodology}
\label{sec:methodology}
\subsection{Data Selection}
We selected four popular GitHub repositories, with each repository containing at least 50K stars. The repositories are: Flutter/flutter, Webpack/webpack, Microsoft/TypeScript, and Angular/angular.
From each repository, we collected the last 10K comments (until 11 November, 2021) from pull requests and issues (5K pull request comments and 5K issue comments). 

%During preprocessing, we replace with placeholders (e.g., $<$URL$>$), remove the usernames, URLs, codes, stracktraces, and quotes (e.g., replying threads, continuation of discussion). 

\subsection{Preprocessing and Dataset Creation}
%To ensure correct emotion detection, we removed noise from the text by applying some preliminary data preprocessing steps.
We preprocessed each issue and pull request comment to replace the url, user mentions, and code with ‘$<$url$>$’, ‘$<$username$>$’, and ‘$<$code$>$’ respectively. 
Consistent with previous research~\cite{Senti4SD, sanei2021impacts}, we did not remove stop words. We also did not include any additional preprocessing, since the emotion classification tools we use (ESEM-E~\cite{esem-e}, EMTk~\cite{emtk}, SEntiMoji~\cite{sentimoji}) have their own preprocessing steps.

%\textcolor{blue}{Please add other details on preprocessing such as "EMTk and SEntiMoji also had their own preprocessing while training, and for ESEM-E, I had added stemming according to their paper." Use this paper for reference “The Impacts of Sentiments and Tones in Community-Generated Issue Discussions”.}

In a preliminary analysis, we observed that many instances in the GitHub comments are neutral, i.e., do not contain any emotion~\cite{Murgia2014DoDF}. Our goal is to avoid creating a sparse dataset with mostly neutral instances, which would be inefficient and time-consuming to annotate. Novielli et al. also made a similar observation and performed selective sampling in creating their dataset of StackOverflow comments~\cite{novielli2018gold}. Hence, to avoid including too many neutral instances, we applied a software engineering-specific sentiment analysis tool~\cite{eeshita-sentiment} to label the preprocessed instances into `positive', `negative', and `neutral'. For each of the four repositories, we randomly selected 250 pull request comments (125 `positive' comments and 125 `negative' comments), 250 issue comments (125 `positive' comments and 125 `negative' comments), to reach a total of 2000 instances.

% first, create a dataset that contains equal number of positive and negative emotions, and second, decreasing manual labor time annotators by labeling neutral utterances.

\subsection{Emotion Categories}
%\textcolor{blue}{please add why, and what do we mean by utterance? An entire PR comment? We need to define it first}. 
As our primary emotion model, we use Shaver's framework~\cite{Shaver} of emotion which has been commonly used in several software engineering studies~\cite{emtk, esem-e}. As shown in Table \ref{tab:shavers_category}, Shaver's framework is a hierarchical (tree-structured) emotion representation model. \updatetext{There are three levels.}
%\textcolor{blue}{In the table, why some tertiary categories have ..? If you think we don't have enough space for that, please mention it here.}
At the top level, there are 6 basic emotion categories: Anger, Love, Fear, Joy, Sadness, and Surprise. For each of the basic emotions, there are secondary and tertiary-level emotions, which refine the granularity of the previous level. For example, Optimism and Hope are the secondary and tertiary level emotions for Joy, respectively.

\begin{table}
\centering
\small
\caption{Shaver's~\cite{Shaver} tree-structured emotion categories \newline (additional categories from GoEmotions~\cite{goemotions} shown in blue)}
\begin{tabular} { | p{1cm} | p{1.4cm} | p{4.8cm} | }
\hline
    Basic Emotion & Secondary Emotion & Tertiary Emotion \\ \hline\hline
    
     & Irritation & Annoyance, Agitation, Grumpiness, Aggravation, Grouchiness \\ \cline{2-3}
     & Exasperation & Frustration \\ \cline{2-3}
     & Rage & Anger, Fury, Hate, Dislike, Resentment, Outrage, Wrath, Hostility, Bitterness, Ferocity, Loathing, Scorn, Spite, Vengefulness \\ \cline{2-3}
     Anger & Envy & Jealousy \\ \cline{2-3}
     & Disgust & Revulsion, Contempt, Loathing \\  \cline{2-3}
     & Torment & - \\ \cline{2-3}
     & \textcolor{blue}{Disapproval} & - \\ \hline
     \hline

     & Affection  & Liking, Caring, Compassion, Fondness, Affection, Love, Attraction, Tenderness, Sentimentality, Adoration \\ \cline{2-3}
    Love & Lust & Desire, Passion, Infatuation \\ \cline{2-3}
     & Longing  & - \\ \hline
    \hline
    & Horror & Alarm, Fright, Panic, Terror, Fear, Hysteria, Shock, Mortification \\ \cline{2-3}
    Fear & Nervousness & Anxiety, Distress, Worry, Uneasiness, Tenseness, Apprehension, Dread \\ \hline
    \hline
     & Cheerfulness & Happiness, Amusement, Satisfaction, Bliss, Gaiety, Glee, Jolliness, Joviality, Joy, Delight, Enjoyment, Gladness, Jubilation, Elation, Ecstasy, Euphoria \\ \cline{2-3}
     & Zest & Enthusiasm, Excitement, Thrill, Zeal, Exhilaration \\ \cline{2-3}
     & Contentment & Pleasure \\ \cline{2-3}
     & Optimism & Eagerness, Hope \\  \cline{2-3}
     Joy & Pride & Triumph \\ \cline{2-3}
     & Enthrallment & Enthrallment, Rapture \\ \cline{2-3}
     & Relief & - \\ \cline{2-3}
     & \textcolor{blue}{Approval} & - \\ \cline{2-3}
     & \textcolor{blue}{Admiration} & - \\ \cline{2-3}
     \hline
    \hline
    & Suffering & Hurt, Anguish, Agony \\ \cline{2-3}
    & Sadness & Depression, Sorrow, Despair, Gloom, Hopelessness, Glumness, Unhappiness, Grief, Woe, Misery, Melancholy \\ \cline{2-3}
    & Disappoint & Displeasure, Dismay \\ \cline{2-3}
    Sadness & Shame & Guilt, Regret, Remorse \\ \cline{2-3}
    & Neglect & Embarrassment, Insecurity, Insult, Rejection, Alienation, Isolation, Loneliness, Homesickness, Defeat, Dejection, Humiliation \\ \cline{2-3}
    & Sympathy & Pity \\ \hline 
    \hline
    & Surprise & Amazement, Astonishment \\ \cline{2-3}
    & \textcolor{blue}{Confusion} & -  \\ \cline{2-3}
    Surprise & \textcolor{blue}{Curiosity} & -  \\ \cline{2-3}
    & \textcolor{blue}{Realization} & - \\ \hline 
\end{tabular}
\label{tab:shavers_category}
\end{table}

%We observed that some commonly used emotions~\cite{goemotions} in Natural Language Processing (NLP) tasks, such as `approval', `disapproval', `confusion', `curiosity', etc. , are missing from Shaver's framework. %We also observed that often developer  communications expresses opinion that can be annotated as such emotions. 
%These missing emotion categories could be often observed in developer communications. 
We observed that some commonly expressed emotions in developer communications, such as Approval, Disapproval, Confusion, Curiosity, etc., are missing from Shaver's framework, which was not designed for emotions expressed in text.
For example, the following GitHub comment can be categorized with the emotion Curiosity, \textit{"I'm curious about this - can you give more context on what exactly goes wrong? Perhaps if that causes bugs this should be prohibited instead?"}, but it does not clearly fit into any of Shaver's existing categories. 
Therefore, to accommodate a wider range of emotions observed in our dataset, we use the recent text-based emotion classification framework by Demszky et al., known as GoEmotions~\cite{goemotions}. 
GoEmotions uses 27 emotions to annotate Reddit comments, which are mapped to Ekman's~\cite{ekman} 6 basic emotion categories. Note that, 5 of Shaver's basic emotions (Anger, Fear, Joy, Sadness, and Surprise) are the same as Ekman's basic categories. Ekman's basic category Disgust is a secondary emotion of Anger in Shaver's categories, and Shaver's basic category Love is a subcategory of Joy in Ekman's basic categories. 

To enhance Shaver's categories, we include selected emotions from GoEmotions~\cite{goemotions} (as shown in Table \ref{tab:shavers_category}). We adopted GoEmotions's definitions and mapping only when an emotion is missing and does not conflict with Shaver's framework. The six additional secondary emotion categories that we adopted from GoEmotions~\cite{goemotions} are highlighted in blue in Table~\ref{tab:shavers_category}. 
\updatetext{Note that all the tertiary level emotions we used come directly from Shaver's framework~\cite{Shaver}; we did not add additional tertiary level emotions.}
%Table~\ref{tab:shavers_category} shows the Shaver's framework with adopted GoEmotion's categories. 

% \textcolor{blue}{This is not a good argument, instead we should add an example emotion that is missing in this framework but is observed frequently in developer communications or in the software engineering domain. Therefore, we combined other emotion models such as GoEmotion and Ekman's model.. we need to add the emotion table we used for annotation.}

% GoEmotion's~\cite{goemotions} used 27 emotions to annotate reddit comments where they mapped these 27 emotions to Ekman's~\cite{Ekman1999} 6 basic categories. Ekman's 6 basic emotions are: anger, disgust, fear, joy, sadness, surprise. GoEmotion mapped `realization' as `surprise`. However, Shaver's framework does not have `realization' as an emotion. In order to help annotators to annotate such utterances and avoid confusion, we adopted some of GoEmotion's mappings and definitions.

% \textcolor{blue}{We need the exact instructions for annotation. Please refer to that and add here.}

\subsection{Data Annotation}
Pull request and issue {\em comments} in GitHub usually consist of multiple sentences. While sometimes each sentence may express different emotions, more often, the comment as a whole conveys an unique emotion. Therefore, in this study, we consider comment-level granularity for data annotation.

Two human judges (authors of this paper) were provided a set of GitHub comments with annotation instructions as follows: \newline
{\em ``You will use the coding schema reported in Table \ref{tab:shavers_category}. For each row in the spreadsheet, please indicate what emotion it conveys (if any) among the basic emotions (first column in the table), which are: Anger, Love, Fear, Joy, Sadness, and Surprise. Multiple emotion labels from the basic emotions are allowed but you should try to avoid them if possible. You can use the second and third levels in the schema as a reference for choosing the primary emotion, but the annotation should be only for the primary emotions. Please mention the second and third-level emotions whenever they are prevalent, and provide a rationale for each annotation. Make sure you consider the emotion(s) of the entire comment and not of individual sentences.''}. The annotation instructions contained more details of the schema including definitions and examples for the basic six emotions. 

\begin{figure}
\centering
\includegraphics[width=1.0\linewidth]{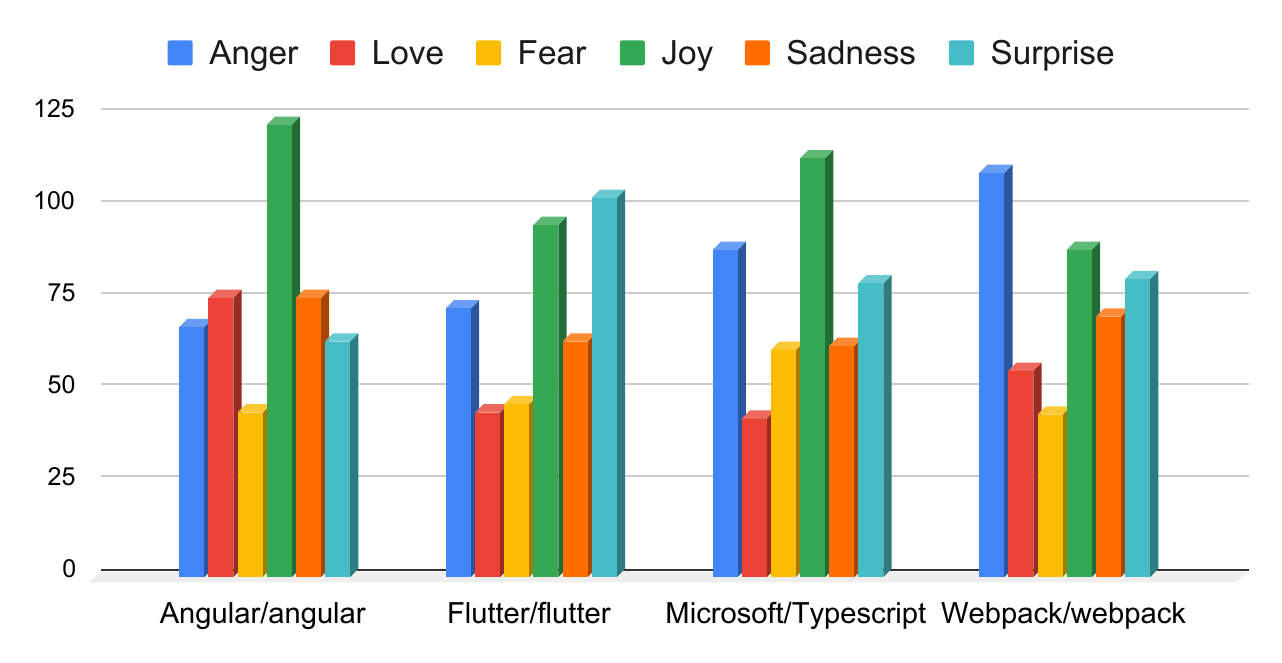}
\caption{Frequency of emotions per project.}
\label{fig:distribution}
\end{figure}

The judges initially annotated a shared set of 400 comments. The  sample size of 400 is sufficient to compute the annotator agreement measure with high confidence~\cite{bujang2017simplified}. The two annotators manually labeled the comments and measured Cohen’s Kappa inter-rater agreement for the six basic emotions. For each of the emotions, they found agreement greater than 0.8, which is considered to be sufficient ($>$ 0.6)~\cite{stemler2004comparison}. The annotators further discussed their annotations until all disagreements were resolved. Afterwards, the annotators separately annotated 800 instances each to reach a total of 2000 utterances. Figure~\ref{fig:distribution} shows the distribution of basic emotion categories per project in the final annotated set. In total, our dataset consists of 340 instances of Anger, 220 instances of Love, 198 instances of Fear, 422 instances of Joy, 274 instances of Sadness, and 328 instances of Surprise.

%{\href{https://docs.google.com/document/d/1pnlNkyjdnbPKrXCutpHWbssjl1dsLugSZx1eC1TEbFY}{annotation instructions}}. 
%We noted that they can use the second and third levels in the schema as a reference for choosing the primary emotion, and asked them to provide the second and third-level emotions when they are identified. We asked them to provide rationale for each selection. Note that, each basic emotion also exists in second or third level emotion. So we asked the annotators whenever they choose just basic emotion, make sure to choose the respective second or third-level emotion. For example, if an instance is labeled \textit{love}, they should select second-level emotion as \textit{affection}, and third-level emotion as \textit{love}. We noted that while multiple emotion labels are allowed, they should try to avoid them if possible. We also emphasised that they should be considering the emotions of the entire unit and not of each sentence belonging to a data point. 

% \begin{figure}
% \centering
% \includegraphics[width=1.0\linewidth]{Images/distribution.png}
% \caption{Emotion distribution with Secondary emotions on the annotated dataset.}
% \label{fig:distribution_alternative}
% \end{figure}

\subsection{Studied Emotion Classification Tools} We investigate three existing software engineering-specific emotion classification tools, which we describe as follows:

\textbf{ESEM-E~\cite{esem-e}}: Murgia et al. proposed an emotion classification tool, which was later referred to as ESEM-E. ESEM-E used Parrott's emotion categories as classification targets~\cite{parrott2001emotions}, which are also featured in Shaver et al.'s model~\cite{Shaver}. While the source code of ESEM-E is not publicly available, we carefully read the descriptions detailed in the paper and implemented it. ESEM-E uses unigram and bigram features and machine learning (ML) models such as SVM, Random Forest, KNN, etc. As recommended by the authors, we use the SVM model. 

\textbf{EMTk~\cite{emtk}}: Calefato et al. proposed EMTk (also known as EmoTxt), which was designed to identify developer emotions from textual communication channels. EMTk identifies six primary emotions according to Shaver's framework \cite{Shaver}. The implemented tool is publicly available on GitHub. EMTk provides two types of data sampling, `NoDownSampling' and `DownSampling'; `DownSampling' randomly samples the majority class to balance the amount of instances between the majority and minority class, while `NoDownSampling' does not change the training data. We use `NoDownSampling' to ensure that all of the three tools use the same training data set.

\textbf{SEntiMoji~\cite{sentimoji}}: Chen et al. proposed SEntiMoji, a transfer learning approach for emotion detection in software engineering (SE) text based on emojis. They concluded that SEntiMoji can significantly outperform existing emotion detection methods (e.g., DEVA~\cite{Deva}, EMTk~\cite{emtk}, MarValous~\cite{MarValous}, ESEM-E~\cite{esem-e}) in software engineering. The SEntiMoji source code is publicly available in GitHub. SEntiMoji is developed based on DeepMoji~\cite{deepmoji} which is an existing deep learning based emoji representation model. \updatetext{The SEntiMoji model can identify various different emotion categorization schemas including Shaver's framework~\cite{Shaver}.}

%When training and testing with each model, we do preprocessing as the respective paper suggested.

% \textbf{GoEmotion \cite{goemotions}}: Demszky et al. released GoEmotions, a manually annotated dataset of 58k English Reddit comments, labeled for 27 emotion categories. This dataset is shown to generalize to other domains and different emotion categories. They also proposed a BERT model that achieves an average F1-score of 0.46 for classifying emotions in their proposed taxonomy.

\subsection{Metrics} We choose popular metrics used to evaluate a classification task: {\em Precision}, {\em Recall} and {F1-score}, which aggregates the prior two. In places where we present combined results across all emotions, we use the micro-averaged variants each of the metrics, as they are responsive to the frequency of occurrence of each constituent emotion.

\begin{itemize}
    \item Precision: Precision is the ratio of true positive observations to the total predicted positive observations. 
    \[Precision = \frac{True Positive}{True Positive + False Positive}\]
    \item Recall: Recall is the ratio of true positive observations to all observations in the class yes. 
    \[Recall = \frac{True Positive}{True Positive + False Negative}\]
    \item F1-score: F1-score is the weighted average of Precision and Recall.
    \[F1-score = 2*\frac{Recall*Precision}{Recall + Precision}\]
\end{itemize}

\subsection{Experiment Design}

Using random stratified sampling~\cite{botev2017variance} for each basic emotion, we divide our annotated dataset into train (80\%) and test (20\%) set. The test set contains a total of 400 data points including 68 instances of Anger, 44 instances of Love, 40 instances of Fear, 84 instances of Joy, 55 instances of Sadness and 65 instances of Surprise.

\begin{table}
\centering
\small
\caption{Comparison of emotion detection tools on GitHub data.}
\begin{tabular} { |c|c|c|c|c|}
\hline
    Emotion & Model & \textbf{Precision} & \textbf{Recall} & \textbf{F1-score}
    % \textbf{Macro-Precision} & \textbf{Macro-Recall} & \textbf{Macro-F1}
    \\ \hline\hline

    & ESEM-E & 0.405  & 0.250  & 0.309 \\ 
    Anger & EMTk & 0.571 & 0.118 & 0.200 \\
    & SEntiMoji & 0.600 & 0.265 & \textbf{0.367} \\
    \hline

    & ESEM-E & 0.651 & 0.636 & \textbf{0.644} \\  
    Love & EMTk & 0.786 & 0.500 & 0.611   \\
    & SEntiMoji & 0.733 & 0.500 & 0.595 \\
    \hline

    & ESEM-E & 0.533 & 0.200 & 0.291 \\  
    Fear & EMTk & 1.00 & 0.200 & \textbf{0.333} \\
    & SEntiMoji & 0.714 & 0.125 & 0.213  \\
    \hline 

    & ESEM-E & 0.458 & 0.321 & \textbf{0.378} \\  
    Joy & EMTk & 0.640 & 0.190 & 0.294  \\
    & SEntiMoji & 0.609 & 0.167 & 0.262 \\
    \hline

    & ESEM-E & 0.759 & 0.400 & \textbf{0.524} \\  
    Sadness & EMTk & 0.778 & 0.382 & 0.512  \\
    & SEntiMoji & 0.857 & 0.327 & 0.474  \\
    \hline
    
    & ESEM-E & 0.596 & 0.431 & 0.500  \\  
    Surprise & EMTk & 0.823 & 0.446 & \textbf{0.580} \\
    & SEntiMoji & 0.846 & 0.338 & 0.484 \\
    \hline \hline

    & ESEM-E & 0.553 & 0.365 & \textbf{0.440} \\  
    {\bf \em Overall} & EMTk & 0.759 & 0.292 & 0.422 \\
    & SEntiMoji & 0.723 & 0.278 & 0.402 \\  \cline{2-5}

    & \cellcolor{l-gray} {\em Average} & \cellcolor{l-gray}0.678 & \cellcolor{l-gray}0.312 & \cellcolor{l-gray}0.421 \\

    \hline
\end{tabular}
\label{tab:user_studies}
\end{table}

% \section{Results and Discussion}

% Our evaluation centers on the two RQs. For the first RQ, we describe the performance of the three SE-specific emotion recognition tools that we study, including an analysis on the types of instances that all tools struggle with. For the second RQ, we describe a data augmentation strategy specific to the task of emotion recognition and discuss its effectiveness.

\section{RQ1: Existing emotion classifiers}

%\textcolor{blue}{How effective are existing emotion classifiers in detecting emotions in GitHub comments? What are common themes?}
\smallskip 
%\begin{mdframed}[backgroundcolor=lightgray!40,topline=false,leftline=false,rightline=false,bottomline=false]
{\bf RQ1:} {\em How effective are existing emotion classifiers in detecting emotions in GitHub comments? What types of emotions are most likely to be misclassified?}
%\end{mdframed}

\noindent
\subsection{Classification Results}
To answer this RQ, we evaluate three well-known tools for software engineering emotion classification (ESEM-E~\cite{esem-e}, EMTk\cite{emtk}, SEntiMoji\cite{sentimoji}), on  our dataset of GitHub issue and pull request discussions (described in Section \ref{sec:methodology}).  
The per-emotion performance of the emotion detection tools is summarized in Table~\ref{tab:user_studies}. The overall trend among all tools is for precision to be significantly higher than recall. In other words, the tools are acting conservatively, choosing to predict more utterances as negative (lacking a certain emotion) than positive, which leads to lower recall. Based on the aggregate measure, F1-score, ESEM-E performed best for Love, Joy and Sadness, EMTk performed best for Fear and Surprise, and SEntiMoji performed best for Anger. 

Results across all emotions are summarized in the bottom part of Table~\ref{tab:user_studies}. Here, on the micro-averaged F1-score metric, ESEM-E improves over the next best tool EMTk by 0.018 (4.3\%). On micro-averaged precision, EMTk improves over SEntiMoji by 0.036 (5.0\%), while on micro-averaged recall, ESEM-E improves the next best tool EMTk by 0.073 (25.0\%). While ESEM-E performs best by far on recall, its performance on precision is worse than EMTk by 0.206 (37.3\%) and SEntiMoji by 0.17 (30.7\%). Overall, across three tools, the average micro F1-score is 0.421.
%the 2nd best tool SEntiMoji by 0.17 (30.74\%).

\begin{figure}[tp]% ! doesn't do what you think it does
\centering
\begin{subfigure}[t]{0.488\linewidth}
    \centering
    %\framebox{
    \includegraphics[width=\linewidth]{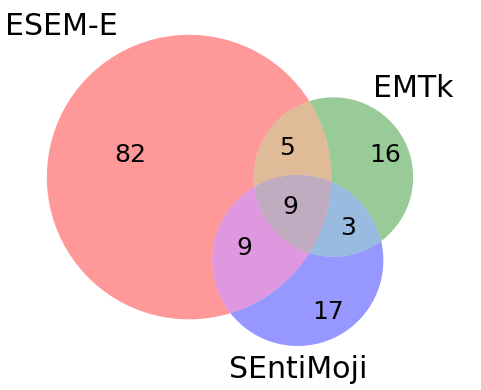}
    %}
    \caption{False positive.}
\end{subfigure}
\begin{subfigure}[t]{0.4\linewidth}
    \centering
    %\framebox{
    \includegraphics[width=\linewidth]{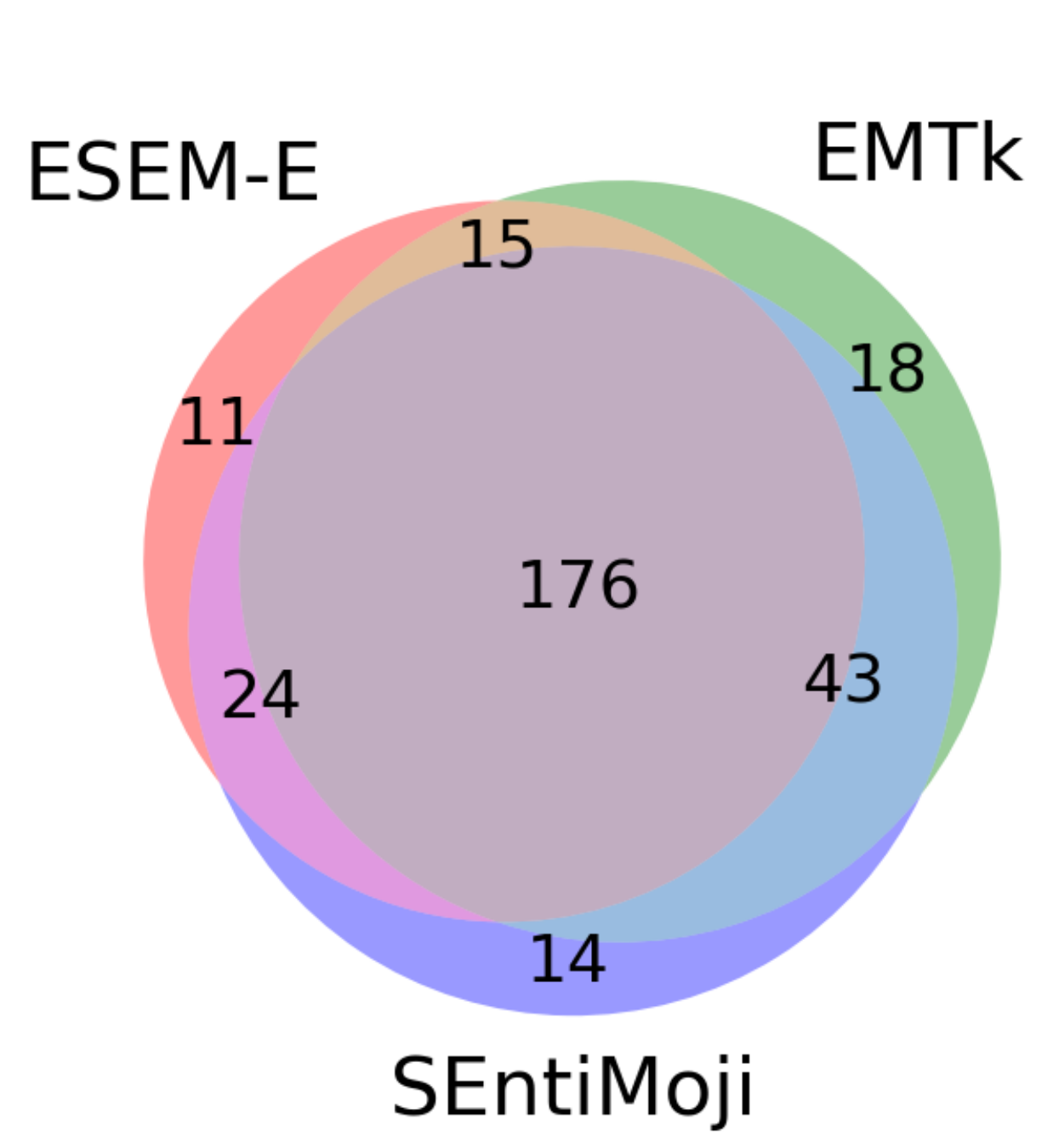}
    %}
    \caption{False negative.}
\end{subfigure}\hfil% equal to outside spacing

\caption{Distribution of FPs and FNs across different tools.}
\label{fig:fpfn}
\end{figure}

% From table~\ref{tab:user_studies}, on f1-score metric, the emotions that show most consistency are love, surprise and sadness; the emotions that show most variability among tools are anger, joy and fear. 
To examine whether the tools tend to struggle on the same instances or if they have complementary strengths, we plot Venn diagrams of the false positive and false negative instances in Figure~\ref{fig:fpfn}. While the false positive instances seem to be broadly spread across different tools, the vast majority of false negatives instances are shared among the three tools. In total, 176/301 (58\%) false negative instances across all emotions were misclassified by all three tools.

\begin{figure}[t]
\centering
\small
\includegraphics[width=1.0\linewidth]{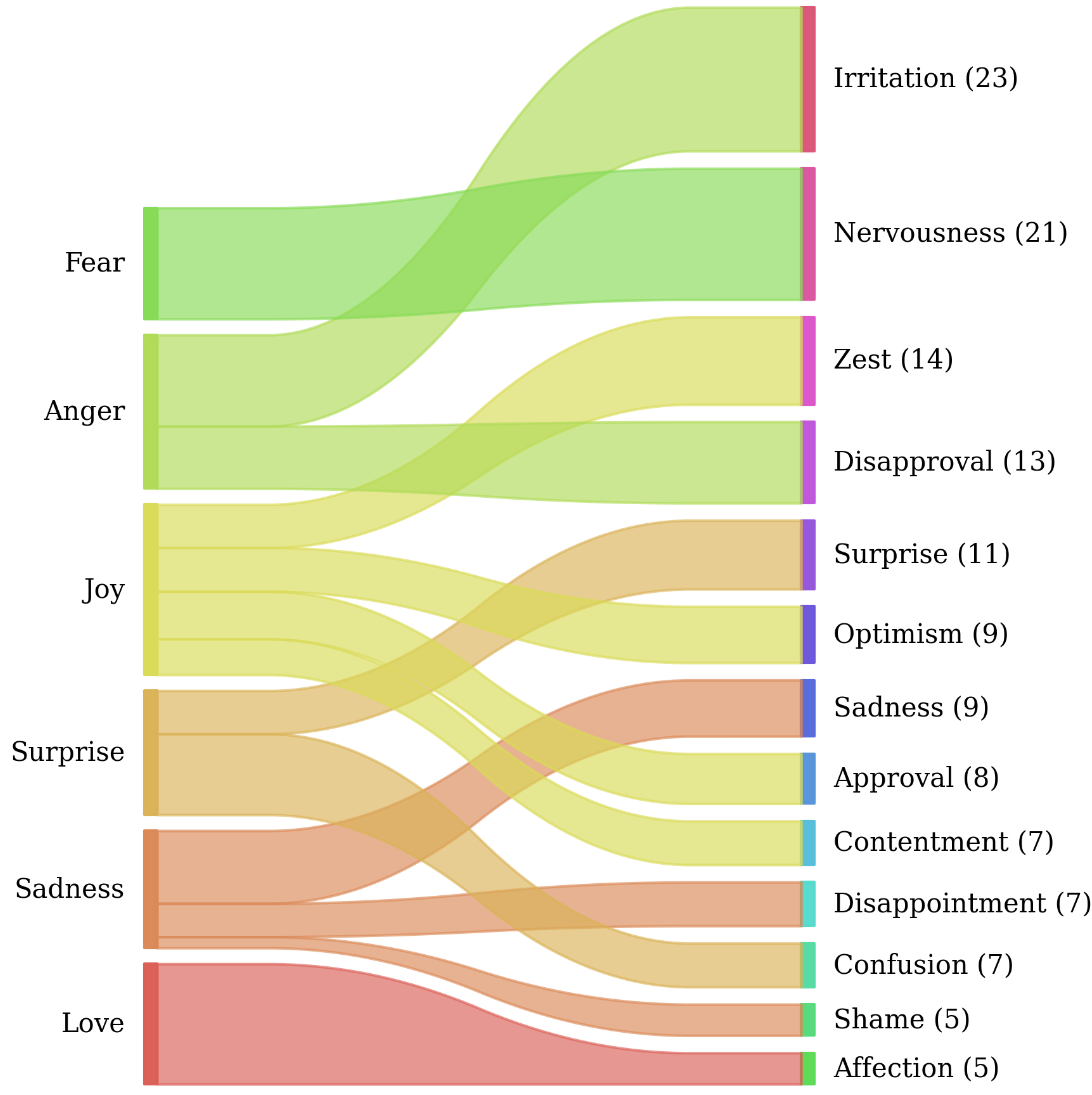}
\caption{FNs mapped to their secondary emotions ($n >= 5$).}
\label{fig:basic_fn}
\end{figure}

\subsection{Error Analysis of FNs}

Because of the unusually high agreement between the tools on the false negative (FN) instances, we focus our error analysis there, i.e., on the 176 FN instances. 

First, we examine the secondary emotions (as listed in Table~\ref{tab:shavers_category}) that are present in the FNs with the goal of understanding if specific emotions are particularly difficult to classify. We create a visualization in Figure~\ref{fig:basic_fn} to understand the distribution of FN instances, i.e., to create the mapping of secondary emotion categories (right side of image) to the six basic level emotions (left side of image) in the FN instances. In this visualization, we only consider secondary emotion categories which has at least 5 FNs. 
The width of the ribbons of top level emotions represent their proportions in the dataset, while the width of the ribbons on the right side represent their proportion of FNs. We observe that some secondary emotions like Irritation, Nervousness and Zest have a significant number of FNs and represent the majority of FNs for basic emotions like Anger, Fear and Joy. For instance, Irritation expressed via comments like \textit{"oh my god, explanation of official documents waste eight hours of me. Why isn't there a case to explain this"}, was misclassified (as FN) 23 out of the 34 times (67.6\%) it appeared in the test set. Nervousness, e.g., \textit{"I guess my concern is that it sets a precedent where somebody could see it and think that it would be fine to use in core"}, was also difficult to recognize as it was misclassified 21 out of 32 times (65.6\%), while Zest was also mistaken often, with a misclassification frequency of 14/18 (77.8\%). Relative to these hard to recognize emotions, Affection, which is part of the Love basic emotion, was a FN only 5 out of 35 times (14.3\%).

% investigate which second level emotions the tools are failing the identify correctly, for the 176 false negative instances shared by all tools, we considered how second level false negative emotions are distributed compared to total basic emotions (). We consider 13 second level emotions that have at least 5 instances in the false negative set. 

% For love and fear, there have only one second level category that have at least 5 counts. It is no surprising that second level emotion affection performed best as X shows that love has least number of false negatives. For fear, its second level emotion nervousness while does not perform reasonably well but does not perform badly either comparing to many other second level emotions. In case of basic emotion joy, the second level emotions varied a lot. While second level emotions cheerfulness and approval performed reasonably well, second level emotion zest and relief performed badly. Similar to joy, in case of anger, sadness and surprise, some second level emotions performed well and some second level emotions did not perform. For anger, second level emotions exasperation and rage performed well but irritation and disapproval do not. For sadness, second level neglect performs well but disappointment does not. For surprise, second level emotions curiosity and confusion perform well but surprise does not. 

Second, to understand the specific difficulties that the tools encountered, we performed a manual qualitative analysis of the 176 FNs. To perform this analysis, we use the error categories defined by Novielli et al.~\cite{novielli2018benchmark} to understand  sentiment classification errors in software engineering text. 

For each of  the FN instances in our dataset, one of authors of this paper performed the initial error mapping, while another author reviewed it and indicated disagreements that were resolved via a discussion. In Table~\ref{tab:error_categories}, we report the distribution of error categories assigned to our FN instances. During the analysis, we observed that multiple error categories can be assigned to some of the FN instances in our dataset. Hence, we chose more than one (i.e., two) error categories for 16 FN instances, while the rest 160 (176 - 16) instances were assigned one error category each.

The most frequent error category found in the FNs is \textit{General error}, which indicates an inability to recognize lexical cues that occur in the text. For instance, in the comment \textit{"that's awesome, I've been needing this for a while"}, annotated as Joy, the tools likely missed clear lexical cues (e.g., the word "awesome"). Similarly for \textit{"oh my god, explanation of official documents waste eight hours of me, Why isn't there a case to explain this"}, annotated as Anger, the tools likely missed the idiomatic expression "waste N hours of me". In other cases, the classifiers miss due to misspellings or broken syntax, as in \textit{"It's anoying me specifically when I want to set it as default value in constructors"}, which is annotated as Anger. \textit{General error} is also the most prevalent in 10 of the 13 secondary categories that are shown in Figure~\ref{fig:basic_fn}.

In 61 cases, the tools failed because of the presence of \textit{Implicit sentiment polarity} in texts. Often, humans use common knowledge to recognize emotions that the tools miss. Consider the following example \textit{``Specifically, I'm less confident in the second commit than the first. AFAICT, it could only return true if a recursive call to itself returned true and all of the recursive base cases returned false."}. This was annotated as Fear (annotators perceived it as an expression of Worry -- a 3rd level emotion that maps to Fear, see Table \ref{tab:shavers_category}), but the emotion is not present explicitly. Sometimes, annotators inferred an emotion based on external knowledge. For instance, \textit{``In that case I would advise you to please file a separate issue with the exact steps and logs to reproduce the issue. Because this issue is about existing apps. Thanks"}, was annotated as Anger since the speaker is expressing a violation of the community rules (Irritation as the secondary emotion). \textit{Implicit sentiment polarity} is the most prevalent in 4 our of the 13 secondary emotions in Figure~\ref{fig:basic_fn} (one category was a tie with \textit{General error}).
As reported by Novielli et al.~\cite{novielli2021assessment}, hostile attitude is often implicit and indirect, which we observe in the error for  Anger's secondary emotions like Exasperation and Irritation. Demszky et al. ~\cite{goemotions} noted that the Nervousness emotion is likely expressed implicitly, which we also observe in our data; the top error category of comments marked with Nervousness is {\em Implicit sentiment polarity}.

% In case of zest, \textit{implicit sentiment polarity} is equally prevalent with \textit{general error}, and for neglect with \textit{figurative language} which is described below. 
%  We also observe that sometimes, some of the second level joy categories shows this kind of error. For example, the following utterance is marked as relief: \textit{This was actually causing this test-case not to be executed!}. As noted b

In 15 cases, we notice that the tools were not able to correctly classify utterances due to \textit{Pragmatics}. This type of error occurs when the annotators consider the context of the comment. In the comment \textit{"hmm, even after a push I still see this test on github, but not locally"}, the author seems to have encountered something unexpected, which the annotators marked as Surprise. 

Sometimes, the use of \textit{Figurative language}, such as  humor, irony, sarcasm, or metaphors, causes difficulties for the classifiers to identify emotions correctly. Often this type of utterances use neutral words to express an emotion. For example, \textit{"Well, if you tried it you'd know"}. In other cases, the lexical presence of \textit{Politeness}, such as "thank you", "please", etc., may cause misclassification. For instance, consider the following example, \textit{"Hi, thanks for your contribution, but we can't review this because you didn’t follow the contributing instructions [...]"}, which is marked as Anger due to the violation of community rules (secondary emotion - Irritation). In other cases, the utterances involve \textit{Polar facts}, that is the utterance invokes an emotion for most people, i.e., the annotators consider the reported situation to invoke an emotion. For instance, in \textit{"I blame the autoformatter."}, the annotators marked this comment as Anger as they considered the author was irritated for facing same problem (second level emotion - Irritation). Overall, we observe that \textit{Figurative language}, \textit{Politeness} and \textit{Polar facts} usually occur in negative emotions (i.e., Anger, Sadness, Fear). The annotation in emotion and sentiment is a subjective task~\cite{scherer2004emotions} as the perception of emotions varies depending on personality trait and personal relevant dispositions. We observe this in 3.1\% cases in our error distribution. \updatetext{
%All 6 cases of these category came from the 1600 utterances that annotators annotated separately. During the error analysis, the annotators disagreed on these cases. 
For instance, \textit{"Do you understand that it is impossible in some cases or can lead to increase size of bundle?"} was considered as Regret (3rd level Sadness) by one annotator, however, the other annotator considered it "Neutral".}

\vspace{0.2cm}

% The main causes of errors are due to the tools missing lexical cues, the use of neutral polarity words to convey implicit emotions, in negative emotions (anger, fear and sadness) the presence of polite words, polar facts, and the use of figurative language to convey emotions.

\noindent
\textbf{Takeaway:} Towards improving or designing new emotion classification tools, some types of errors could be more difficult to address than others. We hypothesize that tools should be able to improve on the most prevalent category of {\em General error} the most, as the lexical cues can be introduced via better training data, which could then be recognized by the tools. Towards that goal, we next investigate if \textit{Data Augmentation} can be an effective strategy to automatically build better training data. %address the problem of inadequate training data.
%\end{mdframed}

\begin{table}[t]
\small
\caption{Distribution of the error categories (as defined by Novielli et al.~\cite{novielli2018benchmark}) in the FN instances.}
\begin{tabular} {|l|c|}
\hline
    Error category & Count
    \\ \hline\hline
    General error & 77 \\
    Implicit sentiment polarity & 61 \\
    Pragmatics & 15 \\
    Figurative language & 15 \\
    Politeness & 10  \\
    Polar facts & 8 \\
    Subjectivity in annotation & 6 \\ \hline
    
\end{tabular}
\label{tab:error_categories}
\end{table}

\section{RQ2: Data Augmentation}
%\begin{mdframed}[backgroundcolor=lightgray!40,topline=false,leftline=false,rightline=false,bottomline=false]
{\bf RQ2:} {\em Can automatic data augmentation techniques be used to improve the effectiveness of existing emotion classifiers?}
%\end{mdframed}

%\noindent
\subsection{Augmentation Strategies} 
We explore three different data augmentation strategies that target emotion classification in software engineering. For each strategy, we use augmentation operators that transform each instance from the training set into a number of augmented instances, each introducing a slightly different vocabulary or idioms into the training set. In fact, we augment by applying a randomly chosen set of our augmentation operators, one after the other in a "stacked" fashion. In some of the augmentation operators, we  rely on recently-introduced generative techniques that are capable of introducing realistic word spans ~\cite{kumar2020data,gan-model-overview}. 
% Specifically, we use BART~\cite{lewis2019bart} as our generative model. 

Via the different augmentation strategies we propose, we explore unconstrained vs. constrained choices of augmented data.
For instance, we examine software-specific vs. generic choices of words to augment with, and how to ensure the original emotions are preserved (or enhanced) by the augmentation. Specifically, we introduce the following three data augmentation strategies: {\em Unconstrained}, {\em Lexicon-based}, and {\em Polarity-based}.

\textbf{{\em Unconstrained Strategy.}} The unconstrained strategy uses augmentation operators that have been previously shown to be effective in NLP and applies them at a randomly chosen location in the text. Inspired by Kumar et al.'s~\cite{kumar2020data} work where they found that a BART-based~\cite{lewis2019bart} generative model outperformed other strategies, we use BART to create generative augmentation operations such as Word Insertion and Word Substitution. 
%The remaining operations follow another technique from NLP, i.e., Easy Data Augmentation (EDA)~\cite{wei-zou-2019-eda}. 
The Unconstrained Strategy uses the following four operators:
\begin{itemize}
    \item {\em Word Insertion using BART:} We insert a word at any position in the original utterance. 
    \item {\em Word Substitution using BART:} We substitute a word at any position.
    \item {\em Word Deletion:} We randomly delete a word at any position.
    \item {\em Sentence Shuffling:} When an utterance has more than one sentence, we randomly shuffle the sentences. 
\end{itemize}

\textbf{{\em Lexicon-based Strategy.}} We observe that sometimes the Unconstrained Strategy produces utterances that may not preserve the original emotion. Note that one of the primary requirements of data augmentation is label invariance, i.e., for the original label to be preserved through the transformation. To deal with this problem, we leverage a software engineering-specific emotion lexicon~\cite{mantyla2017bootstrapping} in order to validate the augmented words generated through the Unconstrained Strategy. Specifically, for each augmented utterance produced by the Unconstrained Strategy, we check if the augmented words exhibit an emotion and if the word's emotion does not match the original emotion. In that case, we replace the word with a software engineering emotion-specific word that preserves the original emotion of the instance. If an utterance is annotated as Joy, and an augmented word exhibits a different emotion (and not Joy), we replace the word with a word from the Joy category of a software engineering-specific lexicon. For example, the Unconstrained Strategy augments the following utterance, which is annotated as Love, \textit{"This looks good, thanks for clarifying the docs."} to \textit{"This looks worse, thanks for reviewing the docs."}. Here the introduction of the word \textit{"worse"} changes the emotion of the original. However, if \textit{"worse"} is replaced with a Love-specific word, i.e., \textit{"wonderful"}, the text becomes \textit{"This looks wonderful, thanks for reviewing the docs."}, which preserves the original label.

As a lexicon, we use NRC's~\cite{mohammad2013nrc} emotion lexicon combined with the software engineering-specific emotional lexicon from M{\"a}ntyl{\"a} et al.'s work~\cite{mantyla2017bootstrapping}, which contains a total of 428 words. Since M{\"a}ntyl{\"a} et al.'s lexicon is not annotated with Shaver's basic emotion categories, we use NRC's emotion lexicon to map each word from M{\"a}ntyl{\"a} et al.'s lexicon to Shaver's basic categories. For example, M{\"a}ntyl{\"a} et al.'s lexicon contains the word \textit{afraid}, which is also available in the NRC emotion lexicon. Since each word in the NRC lexicon is annotated with a specific category (e.g., the word \textit{afraid} is annotated under the emotion category Fear), we map these words to M{\"a}ntyl{\"a} et al.'s lexicon to get a lexicon that is software engineering-specific and also has associated emotion categories.

Note that, as the NRC emotion lexicon uses Plutchik's~\cite{PLUTCHIK19803} emotion categories which has 8 basic emotions, we make two adjustments. First, their basic emotion Disgust is a subcategory in Shaver's basic emotion Anger. Therefore, we combine NRC's Disgust module with Anger module and use it in our Anger lexicon. Second, Plutchik's categories do not contain Shaver's basic category Love, therefore we use NRC's positive module instead as Love; the positive module contains words with positive polarity.

% {\em target dataset}: Operations performed on the target dataset:
% \begin{itemize}
%     \item Word Insertion using BART: For each operation, we insert a word at any position.
%     \item Word Substitution using BART: We substitute words that do not exhibit sentiment polarity~\cite{ahasanuzzaman2018classifying}. 
%     \item Synonym word Substitution: For each operation, we substitute a word that show sentiment polarity with a synonym that also show has similar polarity. We find synonyms using WordNet.
%     \item Word Deletion: We randomly delete a word if it does not exhibit sentiment polarity.
%     \item Sentences Shuffling: When an utterance has more than one sentence, we randomly shuffle sentences. When we do random shuffling of sentences, we do not perform other operations. 
% \end{itemize}

\textbf{{\em Polarity-based Strategy.}} While the Lexicon-based Strategy removes some of the noise that is introduced by the Unconstrained Strategy, we believe the process can be streamlined and the augmentation quality further improved. For instance, a significant problem with the Lexicon-based Strategy is that it uses a lexicon with a very limited number of words. To overcome this constraint, instead of specific emotions, we focus on increasing the polarity words in the augmented instances. Inspired by GoEmotions'~\cite{goemotions} grouping of emotions with sentiment polarity, we formulate three rules that augmented instances have to follow: 1) preserve (or increase) positive polarity words when the annotated utterance is Love and Joy, 2) preserve (or increase) negative polarity words when the annotated utterance is Anger, Fear and Sadness, and 3) preserve the original utterance polarity when the annotated utterance is Surprise.

%, and 4) no polarity check over neutral utterances. 
To identify words that exhibit positive polarity, negative polarity or no polarity, we use %NLTK's~\cite{bird2009natural} Part-Of-Speech (POS) tagger and 
SentiWordNet 3.0~\cite{baccianella2010sentiwordnet}. While ensuring that each valid instance follows the above criteria, we generate new augmented instances using the same operators as for the Unconstrained Dataset. The only modification is that for {\em Word Deletion}, we only randomly delete a word if it does not exhibit sentiment polarity.

% \begin{figure}
% \centering
% \small
% \includegraphics[width=1.0\linewidth]{Images/augmentation_strategies.pdf}
% \caption{Overview of the augmentation strategies.}
% \label{fig:augment_strategies}
% \end{figure}

% \updatetext{
% Figure~\ref{fig:augment_strategies} shows the overview of the three augmentation strategies.
% }

%

% \begin{itemize}
%     \item {\em Word Insertion using BART:} We insert a word at any position.
%     \item {\em Word Substitution using BART:} We substitute a word at any position.
%     \item {\em Word Deletion:} We randomly delete a word if it does not exhibit sentiment polarity.
%     \item {\em Sentences Shuffling:} When an utterance has more than one sentence, we randomly shuffle the sentences.
% \end{itemize}

% Operations performed on the lexicon dataset:
% \begin{itemize}
%     \item Word Insertion using BART: For each operation, we insert a word at any position.
%     \item Word Substitution using lexicon: For each operation, we substitute a word using a lexicon dictionary. We use NRC's emotion lexicon to retrieve a word for respective emotion.
%     \item Word Deletion: We randomly delete a word if it does not exhibit sentiment polarity.
%     \item Sentences Shuffling: When an utterance has more than one sentence, we randomly shuffle sentences. When we do random shuffling of sentences, we do not perform other operations.
% \end{itemize}

\subsection{Augmentation Process} For all three of the above data augmentation strategies, for each instance in our training set, we generate 10 augmented instances, which is considered a reasonable augmentation ratio in the literature~\cite{radford2019language}. For each generated instance, if {\em Sentence Shuffling} is used, we only apply it once. We apply \textit{n} augmentation operations to each instance, where \textit{n} = \textit{max(2, 20\% of the length (i.e., number of words in the instance))}~\cite{zhang2021supporting}. We use \textit{nlpaug}~\cite{ma2019nlpaug} for the generative operations and use \textit{bart-base}~\cite{lewis2019bart} as our BART model's weights. To further ensure that the augmented instance does not change the meaning of the original instance, we added an additional quality check where we ensure that the cosine similarity of BERTOverflow~\cite{tabassum2020code} vectors computed from the augmented and original instance are ($>=$ 0.9) apart~\cite{quteineh2020textual}. BERTOverflow is software engineering-specific version of BERT~\cite{devlin2018bert}, pre-trained on the StackOverflow data dump. We load BERTOverflow using the {\em huggingface} library~\cite{huggingface}.

\subsection{Augmentation Results and Discussion} 

% Kosta: Start with Overall

% Table~\ref{tab:augmentation_results}

Overall, across all three tools, all of the augmentation strategies improved performance over the original results (Table~\ref{tab:augmentation_results}).
The average micro F1-score improvement with the Unconstrained Strategy is 4.8\% (0.441), with the Lexicon Strategy is 7.8\% (0.455), and with the Polarity Strategy is 9.3\% (0.461). Considering the three tools separately, we observe improved F1-score across the board, with EMTk benefiting the most using the Polarity Strategy with an improvement in F1-score of 13.7\%. 

The Unconstrained Strategy worked best with SEntiMoji by improving the F1-score by 7.7\%. Both Lexicon Strategy and Polarity Strategy improved most with EMTk by 10.7\% and 13.7\% respectively. The reasons likely lie behind the feature extraction methods of EMTk, as its classification features are based on an emotion lexicon and polarity. 

ESEM-E performed best with the Lexicon Strategy, outperforming the original dataset by 6.8\%. As ESEM-E directly uses unigrams and bigrams as its features, it is likely that the repetition of lexical cues produced by the Lexicon Strategy significantly helped this tool. EMTk performed most effectively with the Polarity Strategy (13.7\% improvement) as positive and negative sentiment polarity scores are one of its features. SEntiMoji performed best with the Polarity Strategy as well, outperforming the original dataset by 8.0\%; however, SEntiMoji's performance did not vary significantly over all three augmentation approaches.

The emotions that are improved most with data augmentation strategies are Sadness and Joy, which is consistent with Murgia et al.'s~\cite{Murgia2014DoDF} findings that they are easier to identify compared to other basic emotions. The reason is likely because data augmentation helped to introduce more lexical cues that were missing in the original dataset. In our analysis for RQ1, we observed that the most prevalent error category for Sadness and Joy FNs was \textit{General Error}. Sadness achieved maximum F1-score of 0.557 in Unconstrained Strategy with SEntiMoji, and Joy achieved a maximum F1-score of 0.406 with the Polarity Strategy and ESEM-E. 
Surprise performed best with the Polarity Strategy where all three tools improved the F1-score with EMTk producing the best result, an F1-score of 0.630. Previous research shows that Surprise in SE is generally hard to detect, since it is not very frequent~\cite{cabrera2020classifying, opinionlitreview}. However, with the addition of GoEmotions's categories and data augmentation, detection of Surprise improved significantly.
As noted in previous research~\cite{cabrera2020classifying, Murgia2014DoDF}, Anger and Fear are difficult to predict, as they often depend on the message context. During our error analysis, we saw that most \textit{Implicit sentiment polarity} errors occur with Anger and Fear. These type of errors were difficult to identify even after data augmentation. SEntiMoji did not improve Anger performance in any of the strategies; while EMTk improved, its performance with the initial dataset was very low. In the case of Fear, ESEM-E did not improve with any of the strategies. However, Fear performed significantly better with the Polarity Strategy in EMTk, achieving F1-score of 0.473. Further research on what caused EMTk to perform better for Fear may help to pinpoint how to further improve classifying this emotion.

One interesting case is that with the original dataset, Love performed best across all three tools, however, with data augmentation, the performance of Love did not improve significantly. This points to a limitation of data augmentation in that it can only be of a limited benefit, i.e., useful only in cases where sufficient lexical cues are not already present in the data.

\vspace{0.2cm} 
\noindent
\textbf{Takeaway:} Overall, we observe that Data Augmentation generally improves emotion classification performance across different emotions and tools. We observe improvements especially when the initial dataset has insufficient lexical cues for a specific emotion. Out of the three augmentation strategies we experimented with, the Polarity Strategy worked really well, as it provided a balance between completely unconstrained augmentation (which introduces noise) and highly constrained augmentation (which fails to increase size and diversity). Data augmentation is likely only able to improve performance up to point, as our current augmentation operators do not seem to help in identifying implicit emotions, such as Sarcasm.

\begin{table}
\centering
\footnotesize
\caption{Emotion classification results for all three data augmentation strategies. For F1-score, we also show the percentage improvement over the original (unaugmented) dataset.}
%\begin{adjustwidth}{-0.5cm}{}
% \def\arraystretch{1.2}% 
\begin{tabular} { | c | c | c | c | c | c | }
\hline
    Emotion & Strategy & Model & Precision & Recall & F1-score \\ \hline\hline 

    % & & ESEM-E & 0.405  & 0.250  & 0.309 \\ 
    % & NoAug & EMTk & 0.571 & 0.118 & 0.200 \\
    % & & SEntiMoji & 0.600 & 0.265 & 0.367 \\ \cline{2-6}
    & & ESEM-E & 0.567 & 0.250 & 0.347 (12.3\%) \\ 
    & Unconstrained & EMTk & 0.571 & 0.235 & 0.333 (66.5\%) \\ 
    & & SEntiMoji & 0.630 & 0.250 & \textbf{0.358} (-2.5\%) \\ \cline{2-6}
    & & ESEM-E & 0.581 & 0.265 & 0.364 (17.8\%) \\ 
    Anger & Lexicon & EMTk & 0.531 & 0.250 & 0.340 (70.0\%) \\ 
    & & SEntiMoji & 0.625 & 0.221 & 0.326 (-11.2\%) \\  \cline{2-6}
    & & ESEM-E & 0.500 & 0.235 & 0.320 (3.6\%) \\ 
    & Polarity & EMTk & 0.609 & 0.206 & 0.308 (54.0\%) \\
    & & SEntiMoji & 0.615 & 0.235 & 0.340 (-7.4\%) \\  \cline{2-6}
    \hline  

    % & & ESEM-E & 0.651 & 0.636 & 0.644 \\  
    % & NoAug & EMTk & 0.786 & 0.500 & 0.611   \\
    % & & SEntiMoji & 0.733 & 0.500 & 0.595 \\ \cline{2-6}
    & & ESEM-E & 0.596 & 0.636 & 0.615 (-4.5\%) \\ 
    & Unconstrained & EMTk & 0.703 & 0.591 & 0.642 (5.1\%) \\
    & & SEntiMoji & 0.719 & 0.523 & 0.605 (1.7\%)  \\ \cline{2-6}
    & & ESEM-E & 0.630 & 0.659 & 0.644 (0.0\%) \\ 
    Love & Lexicon & EMTk & 0.659 & 0.614 & 0.635 (3.9\%) \\ 
    & & SEntiMoji & 0.710 & 0.500 & 0.587 (-1.3\%) \\  \cline{2-6}
    & & ESEM-E & 0.667 & 0.682 & \textbf{0.674} (4.7\%) \\ 
    & Polarity & EMTk & 0.727 & 0.545 & 0.623 (2.0\%) \\ 
    & & SEntiMoji & 0.733 & 0.500 & 0.595 (0.0\%) \\  \cline{2-6}
    \hline  

    % & & ESEM-E & 0.533 & 0.200 & 0.291 \\  
    % & NoAug & EMTk & \textbf{1.00} & 0.200 & 0.333 \\
    % & & SEntiMoji & 0.714 & 0.125 & 0.213  \\ \cline{2-6}
    & & ESEM-E & 0.545 & 0.150 & 0.235 (-19.2\%) \\ 
    & Unconstrained & EMTk & 0.600 & 0.225 & 0.327 (-1.8\%) \\
    & & SEntiMoji & 0.700 & 0.175 & 0.280 (31.5\%) \\ \cline{2-6}
    & & ESEM-E & 0.600 & 0.150 & 0.231 (-20.6\%) \\ 
    Fear & Lexicon & EMTk & 0.818 & 0.225 & 0.353 (6.0\%) \\ 
    & & SEntiMoji & 0.636 & 0.175 & 0.275 (29.1\%) \\  \cline{2-6}
    & & ESEM-E & 0.500 & 0.150 & 0.231 (-20.6\%) \\ 
    & Polarity & EMTk & 0.867 & 0.325 & \textbf{0.473} (42.0\%) \\ 
    & & SEntiMoji & 0.600 & 0.150 & 0.240 (12.7\%) \\  \cline{2-6}
    \hline  

    % & & ESEM-E & 0.458 & 0.321 & 0.378  \\  
    % & NoAug & EMTk & 0.640 & 0.190 & 0.294  \\
    % & & SEntiMoji & 0.609 & 0.167 & 0.262 \\ \cline{2-6}
    & & ESEM-E & 0.456 & 0.310 & 0.369 (-2.4\%) \\ 
    & Unconstrained & EMTk & 0.486 & 0.214 & 0.298 (1.4\%) \\ 
    & & SEntiMoji & 0.477 & 0.250 & 0.328 (25.2\%) \\ \cline{2-6}
    & & ESEM-E & 0.500 & 0.321 & 0.391 (3.4\%) \\
    Joy & Lexicon & EMTk & 0.590 & 0.274 & 0.374 (27.2\%) \\
    & & SEntiMoji & 0.526 & 0.238 & 0.328 (25.2\%) \\  \cline{2-6}
    & & ESEM-E & 0.492 & 0.345 & \textbf{0.406} (7.4\%) \\ 
    & Polarity & EMTk & 0.613 & 0.226 & 0.330 (12.2\%) \\ 
    & & SEntiMoji & 0.575 & 0.274 & 0.371 (41.6\%) \\  \cline{2-6}
    \hline  

    % & & ESEM-E & 0.759 & 0.400 & 0.524 \\  
    % & NoAug & EMTk & 0.778 & 0.382 & 0.512  \\
    % & & SEntiMoji & 0.857 & 0.327 & 0.474  \\ \cline{2-6}
    & & ESEM-E & 0.767 & 0.418 & 0.541 (3.2\%) \\ 
    & Unconstrained & EMTk & 0.909 & 0.364 & 0.519 (1.4\%) \\ 
    & & SEntiMoji & 0.917 & 0.400 & \textbf{0.557} (17.5\%) \\ \cline{2-6}
    & & ESEM-E & 0.759 & 0.400 & 0.524 (0.0\%) \\ 
    Sadness & Lexicon & EMTk & 0.719 & 0.418 & 0.529 (3.3\%) \\ 
    & & SEntiMoji & 0.875 & 0.382 & 0.532 (12.2\%) \\  \cline{2-6}
    & & ESEM-E & 0.710 & 0.400 & 0.512 (-2.3\%) \\
    & Polarity & EMTk & 0.821 & 0.418 & 0.554 (8.2\%) \\ 
    & & SEntiMoji & 0.913 & 0.382 & 0.538 (13.5\%) \\  \cline{2-6}
    \hline  

    % & & ESEM-E & 0.596 & 0.431 & 0.500  \\  
    % & NoAug & EMTk & 0.823 & 0.446 & 0.580   \\
    % & & SEntiMoji & 0.846 & 0.338 & 0.484 \\ \cline{2-6}
    & & ESEM-E & 0.646 & 0.477 & 0.549 (9.8\%) \\ 
    & Unconstrained & EMTk & 0.784 & 0.446 & 0.569 (-1.9\%) \\ 
    & & SEntiMoji & 0.767 & 0.354 & 0.484 (0.0\%) \\ \cline{2-6}
    & & ESEM-E & 0.667 & 0.523 & 0.586 (17.2\%) \\ 
    Surprise & Lexicon & EMTk & 0.732 & 0.462 & 0.566 (-2.4\%) \\ 
    & & SEntiMoji & 0.857 & 0.369 & 0.516 (6.6\%) \\  \cline{2-6}
    & & ESEM-E & 0.654 & 0.523 & 0.581 (16.2\%) \\ 
    & Polarity & EMTk & 0.791 & 0.523 & \textbf{0.630} (8.6\%) \\ 
    & & SEntiMoji & 0.852 & 0.354 & 0.500 (3.3\%) \\  \cline{2-6}
    \hline \hline

    % & & ESEM-E & 0.553 & 0.365 & 0.440 \\  
    % & NoAug & EMTk & 0.759 & 0.292 & 0.422 \\
    % & & SEntiMoji & 0.723 & 0.278 & 0.402 \\ \cline{2-6}
    & & ESEM-E & 0.587 & 0.368 & 0.453 (3.0\%) \\ 
    & Unconstrained & EMTk & 0.659 & 0.326 & 0.436 (3.3\%) \\ 
    & & SEntiMoji & 0.681 & 0.317 & 0.433 (7.7\%) \\  \cline{3-6}
    & & \cellcolor{l-gray}{\em Average} & \cellcolor{l-gray}0.642 & \cellcolor{l-gray}0.337 & \cellcolor{l-gray}0.441 (4.8\%) \\  \cline{2-6}
    & & ESEM-E & 0.610 & 0.382 & 0.470 (6.8\%) \\ 
    {\em \textbf{Overall}} & Lexicon & EMTk & 0.658 & 0.362 & 0.467 (10.7\%) \\ 
    & & SEntiMoji & 0.699 & 0.306 & 0.426 (6.0\%) \\  \cline{3-6}
    & & \cellcolor{l-gray}{\em Average} & \cellcolor{l-gray}0.656 & \cellcolor{l-gray}0.350 & \cellcolor{l-gray}0.454 (7.8\%) \\  \cline{2-6}
    & & ESEM-E & 0.593 & 0.385 & 0.467 (6.1\%) \\ 
    & Polarity & EMTk & 0.734 & 0.357 & \textbf{0.480} (13.7\%) \\
    & & SEntiMoji & 0.712 & 0.312 & 0.434 (8.0\%) \\  \cline{3-6}
    & & \cellcolor{l-gray}{\em Average} & \cellcolor{l-gray}0.680 & \cellcolor{l-gray}0.351 & \cellcolor{l-gray}0.460 (9.3\%) \\  \cline{2-6}

    \hline
\end{tabular}
%\end{adjustwidth}
\label{tab:augmentation_results}
\end{table}

\section{Related Work}
Below, we describe the related work sourced from two different domains: emotion analysis of software artifacts, and data augmentation in the domain of Natural Language Processing (NLP).

\subsection{Emotion Analysis of Software Artifacts}
One of the earliest emotion analysis of software artifacts can be found in Murgia et al.'s work~\cite{Murgia2014DoDF}, where they manually analyzed 800 issue comments and concluded that some basic emotions such as Love, Joy and Sadness are easier to identify in text. In their later work, Murgia et al.~\cite{esem-e} proposed an automated approach (namely, ESEM-E) to detect emotions in software artifacts. Motivated by their earlier work, they only focused on automatically identifying Love, Joy and Sadness in Ortu el al.'s JIRA-based dataset~\cite{ortu2016}. Calefato et al.~\cite{emtk} developed a feature extraction-based machine learning technique EMTk (also known as EmoTxt). They evaluated EMTk on the StackOverflow dataset initially developed by Novielli et. al.~\cite{novielli2018gold}, which was annotated with Shaver's six basic emotions~\cite{Shaver}. Neupane et al.~\cite{emod} investigated emotion dynamics in GitHub repositories, using EMTk as their primary model. Cabrera-Diego et al.~\cite{cabrera2020classifying} used the multi-label classifiers HOMER~\cite{tsoumakas2008effective} and RAkEL~\cite{tsoumakas2010random} on the above mentioned JIRA and StackOverflow datasets, observing that HOMER and RAkEL achieved a better micro F1-score than EMTk on the JIRA dataset and similar performance on the StackOverflow dataset. They concluded that EMTk is a conservative tool in general. Both the JIRA and StackOverflow dataset, however, have a limited number of utterances in certain emotion categories. The JIRA dataset only focuses on four categories of emotions (Love, Sadness, Joy, and Anger), while the StackOverflow dataset has a prevalence of Love compared to other categories and has a very limited number of Surprise instances. One of our focus areas in this paper has been on curating a dataset where all the emotion categories are represented in sufficient quantity.

Venigalla et al.~\cite{venigalla-emotion} analyzed software developers' emotions towards software documentation using the NRC emotion analyzer module~\cite{mohammad2013nrc} of the Syuzhet package~\cite{Syuzhet}. In our work, we also use the NRC's emotion module, but for data augmentation. Another thread of research in this area is based on the VAD (Valence, Arousal, and Dominance) model~\cite{russell1977evidence}, where specific emotions are represented as a mixture of these three numerically-expressed quantities. For instance, VAD can express emotions such as, Excitement (positive valence and high arousal), Relaxation (positive valence and low arousal), Depression (negative valence and low arousal), and Stress (negative valence and high arousal). A tool named DEVA~\cite{Deva} was one of the first in software engineering that was based on the VAD model. Later, researchers proposed MarValous~\cite{MarValous}, which significantly outperformed DEVA (by 23.05\%). Another focus of recent research has been through the use of emojis in various software artifact such as pull requests, issue comments, chats, GitHub discussions, etc.~\cite{son2021more, lu2018first, rong2022empirical, Chatterjee2019, Chatterjee2020Journal}. Chen el al. proposed a technique called SEntiMoji that uses emojis to recognize emotions~\cite{sentimoji, chen2021emoji}, based on an existing emoji representation model DeepMoji~\cite{deepmoji}. SEntiMoji is able to produce output in Shaver's categories or in the VAD model. When SEntiMoji was compared against EMSE-E~\cite{esem-e} and EMTk~\cite{emtk} on the JIRA and StackOverflow dataset, it showed overall improvement across all six basic emotions. SEntiMoji was also compared against DEVA~\cite{Deva} and MarValous~\cite{MarValous}, and was shown to achieve better performance than these two tools as well. In our research, we compare the tools that use Shaver's emotion categories, SEntiMoji, EMSM-E, and EMTk, using a GitHub-based dataset that we curate. We consider data augmentation as the means to improve performance in all three of these tools.

\subsection{Data Augmentation in NLP}
The augmentation of textual data (i.e., in NLP) has been an area of considerable interest in recent years to address the data scarcity problem related to different tasks such as, subjectivity detection~\cite{Guo2019AugmentingDW}, question understanding and summarization~\cite{mrini-etal-2021-gradually}.
%, event causality identification~\cite{zuo-etal-2021-learnda}. 
% detecting positive/negative reviews (Pang and Lee, 2005); a subjectivity detection dataset for classifying a sentence as being subjective or objective (Pang and Lee, 2004).
Early techniques of interest in data augmentation included synonym replacement~\cite{zhang2015character}, i.e., replacing a word with its synonym, and BackTranslation~\cite{backtranslation}, i.e., paraphrasing a text by converting to a second language and then converting back to original language. Inspired by similar approaches in computer vision, researchers also devised MixUp augmentation~\cite{mixup, shorten2021text}, which meshes together existing examples in order to create new augmented instances. Another recent research direction is Unsupervised Data Augmentation~\cite{xie2019unsupervised} (UDA) which is a semi-supervised model that outperformed state-of-the-art methods using only 20 labeled instances. Wei el al.~\cite{wei-zou-2019-eda} proposed a method - Easy Data Augmentation (EDA) which worked surprisingly well for smaller datasets despite being a simple technique that uses straightforward operators, e.g., synonym replacement using WordNet~\cite{miller1995wordnet}, random insertion, random swap, and random deletion. As software artifact datasets are often small~\cite{smalldataset}, we were inspired by EDA in devising our data augmentation technique. 
A recent research thread in data augmentation is contextual augmentation~\cite{shorten2021text}, using various large language models such as CBERT~\cite{cbert}, BART~\cite{kumar2020data}, GPT-2~\cite{anaby2020not}, etc. Kumer at al.~\cite{kumar2020data} showed that for classification tasks, BART outperforms other models due to its ability to generate longer sequences of text in context. In this paper, we also leverage the BART model as part of our augmentation operators.

\section{Threats to Validity} Several limitations may impact the interpretation of our findings. We categorize and list each of them below.

\textbf{\textit{Construct validity}}. Construct validity concerns the relationship between theory and observation. Shaver's emotions model~\cite{Shaver} and the GoEmotions model by Demszky et al.~\cite{goemotions} are two different schemas. We use a combination of the two, which may violate their original design. To mitigate this risk, we carefully read both of the original research and used Shaver's model as our primary schema, integrating GoEmotions' categories only when they are complementary and do not conflict with Shaver's in any part of their definitions. Furthermore, the error analysis in RQ1 shows that none of the emotion categories that integrated GoEmotions are among the worse performing. Instead, the addition of GoEmotions secondary emotion categories specifically improves the performance of the basic category Surprise, which has exhibited a relatively low F1-score in previous research in software engineering emotion classification  ~\cite{opinionlitreview}.

\textbf{\textit{Internal validity}}. Internal validity concerns the study design factors that may influence the results. One such threat to our study is not doing cross-validation, which would have improved the reliability of the results. We mitigate this threat by using stratified sampling and a reasonable train-test data split of 80\%-20\% respectively. Another threat is that, to conduct our experiments we use existing tools and their released code, except for ESEM-E~\cite{esem-e}. %which we implemented based on the authors' provided description. 
%To mitigate that threat, we have followed the authors' instructions to the best of our ability. 
It is possible that we have incorrectly implemented ESEM-E, although, we explicitly followed the authors' instructions to mitigate this threat.
%the design of ESEM-E is fairly straightforward, so we believe this to be unlikely. 
The subjectivity of annotating emotions (and in the error analysis) presents another threat to internal validity. However, the use of a three-tiered emotion structure and high inter-rater agreement ($>$ 0.8) ensures the reliability of the annotation procedure. 

\textbf{\textit{External validity}}. External validity concerns the generalization of our findings. Our study shows that data augmentation improves emotion classification across the three tools we experimented with. However, the specific augmentation strategies may not generalize beyond the three tools we studied and our dataset extracted from GitHub comments. More specifically, our findings may not generalize over other types of artifacts in software engineering, such as StackOverflow, JIRA, etc. While our results introduce the potential of data augmentation for emotion classification, further investigation is needed in other to validate our results beyond the tools and the data used in our study. 

\section{Conclusion and Future Work}
In this paper, we first conduct a qualitative study to understand the limitations of the existing machine learning-based tools for classifying emotions in software engineering-related text. Specifically, we evaluate ESEM-E~\cite{esem-e}, EMTk~\cite{emtk}, SEntiMoji~\cite{sentimoji} on our curated dataset of 2000 GitHub pull requests and issue comments. We observe that some types of errors could be more difficult to address than others, however there is a scope to improve the performance of the existing tools by creating better training data. Thus, next we investigate three types of data augmentation strategies that could be leveraged to improve emotion detection in software-related text. Our results indicate that augmentation operators that target words with specific polarity are significantly more effective than generic augmentation operators. Using polarity-based augmentation shows an average improvement of 9.3\% in micro F1-Score across the three existing emotion classification tools. %This research provides a significant contribution to ...

Our immediate next steps focus on investigating more diverse data sources, including other types of software artifacts such as StackOverflow, Slack chats, etc. We are also interested in exploring new data augmentation techniques based on large language models that are pre-trained on software engineering corpora and fine-tuned to emotion and sentiment-type tasks. Finally, we would like to explore if polarity based data augmentation strategies could improve other related tasks in software engineering, such as sentiment analysis.

\balance 
\bibliographystyle{ACM-Reference-Format}
\bibliography{references}

@inproceedings{sinha2016analyzing,
  title={Analyzing developer sentiment in commit logs},
  author={Sinha, Vinayak and Lazar, Alina and Sharif, Bonita},
  booktitle={Proceedings of the 13th international conference on mining software repositories},
  pages={520--523},
  year={2016}
}

@inproceedings{kumar2020data,
  title={Data Augmentation using Pre-trained Transformer Models},
  author={Kumar, Varun and Choudhary, Ashutosh and Cho, Eunah},
  booktitle={Proceedings of the 2nd Workshop on Life-long Learning for Spoken Language Systems},
  pages={18--26},
  year={2020}
}

@inproceedings{ortu2015bullies,
  title={Are bullies more productive? Empirical study of affectiveness vs. issue fixing time},
  author={Ortu, Marco and Adams, Bram and Destefanis, Giuseppe and Tourani, Parastou and Marchesi, Michele and Tonelli, Roberto},
  booktitle={2015 IEEE/ACM 12th Working Conference on Mining Software Repositories},
  pages={303--313},
  year={2015},
  organization={IEEE}
}

@book{parrott2001emotions,
  title={Emotions in social psychology: Essential readings},
  author={Parrott, W Gerrod},
  year={2001},
  publisher={psychology press}
}

@inproceedings{quteineh2020textual,
  title={Textual data augmentation for efficient active learning on tiny datasets},
  author={Quteineh, Husam and Samothrakis, Spyridon and Sutcliffe, Richard},
  booktitle={Proceedings of the 2020 Conference on Empirical Methods in Natural Language Processing (EMNLP)},
  pages={7400--7410},
  year={2020},
  organization={Association for Computational Linguistics}
}

@incollection{PLUTCHIK19803,
  title={A general psychoevolutionary theory of emotion},
  author={Plutchik, Robert},
  booktitle={Theories of emotion},
  pages={3--33},
  year={1980},
  publisher={Elsevier}
}

@inproceedings{anaby2020not,
  title={Do not have enough data? Deep learning to the rescue!},
  author={Anaby-Tavor, Ateret and Carmeli, Boaz and Goldbraich, Esther and Kantor, Amir and Kour, George and Shlomov, Segev and Tepper, Naama and Zwerdling, Naama},
  booktitle={Proceedings of the AAAI conference on artificial intelligence},
  volume={34},
  pages={7383--7390},
  year={2020}
}

@article{smalldataset,
  title={Making the most of small software engineering datasets with modern machine learning},
  author={Prenner, Julian Aron and Robbes, Romain},
  journal={IEEE Transactions on Software Engineering},
  volume={48},
  number={12},
  pages={5050--5067},
  year={2021},
  publisher={IEEE}
}

@inproceedings{emtk,
  title={Emtk-the emotion mining toolkit},
  author={Calefato, Fabio and Lanubile, Filippo and Novielli, Nicole and Quaranta, Luigi},
  booktitle={2019 IEEE/ACM 4th International Workshop on Emotion Awareness in Software Engineering (SEmotion)},
  pages={34--37},
  year={2019},
  organization={IEEE}
}

@inproceedings{sentimoji,
  title={Sentimoji: an emoji-powered learning approach for sentiment analysis in software engineering},
  author={Chen, Zhenpeng and Cao, Yanbin and Lu, Xuan and Mei, Qiaozhu and Liu, Xuanzhe},
  booktitle={Proceedings of the 2019 27th ACM joint meeting on european software engineering conference and symposium on the foundations of software engineering},
  pages={841--852},
  year={2019}
}

@article{Shaver,
  title={Emotion knowledge: further exploration of a prototype approach.},
  author={Shaver, Phillip and Schwartz, Judith and Kirson, Donald and O'connor, Cary},
  journal={Journal of personality and social psychology},
  volume={52},
  number={6},
  pages={1061},
  year={1987},
  publisher={American Psychological Association}
}

@article{esem-e,
  title={An exploratory qualitative and quantitative analysis of emotions in issue report comments of open source systems},
  author={Murgia, Alessandro and Ortu, Marco and Tourani, Parastou and Adams, Bram and Demeyer, Serge},
  journal={Empirical Software Engineering},
  volume={23},
  number={1},
  pages={521--564},
  year={2018},
  publisher={Springer}
}

@inproceedings{Deva,
  title={DEVA: sensing emotions in the valence arousal space in software engineering text},
  author={Islam, Md Rakibul and Zibran, Minhaz F},
  booktitle={Proceedings of the 33rd annual ACM symposium on applied computing},
  pages={1536--1543},
  year={2018}
}

@inproceedings{MarValous,
  title={Marvalous: Machine learning based detection of emotions in the valence-arousal space in software engineering text},
  author={Islam, Md Rakibul and Ahmmed, Md Kauser and Zibran, Minhaz F},
  booktitle={Proceedings of the 34th ACM/SIGAPP Symposium on Applied Computing},
  pages={1786--1793},
  year={2019}
}

@article{rong2022empirical,
  title={An empirical study of emoji use in software development communication},
  author={Rong, Shiyue and Wang, Weisheng and Mannan, Umme Ayda and de Almeida, Eduardo Santana and Zhou, Shurui and Ahmed, Iftekhar},
  journal={Information and Software Technology},
  pages={106912},
  year={2022},
  publisher={Elsevier}
}

@article{mohammad2013nrc,
  title={Nrc emotion lexicon},
  author={Mohammad, Saif M and Turney, Peter D},
  journal={National Research Council, Canada},
  volume={2},
  year={2013}
}

@inproceedings{novielli2018benchmark,
  title={A benchmark study on sentiment analysis for software engineering research},
  author={Novielli, Nicole and Girardi, Daniela and Lanubile, Filippo},
  booktitle={2018 IEEE/ACM 15th International Conference on Mining Software Repositories (MSR)},
  pages={364--375},
  year={2018},
  organization={IEEE}
}

@inproceedings{mantyla2017bootstrapping,
  title={Bootstrapping a lexicon for emotional arousal in software engineering},
  author={M{\"a}ntyl{\"a}, Mika V and Novielli, Nicole and Lanubile, Filippo and Claes, Ma{\"e}lick and Kuutila, Miikka},
  booktitle={2017 IEEE/ACM 14th International Conference on Mining Software Repositories (MSR)},
  pages={198--202},
  year={2017},
  organization={IEEE}
}

@inproceedings{backtranslation,
  title={Improving neural machine translation models with monolingual data},
  author={Sennrich, Rico and Haddow, Barry and Birch, Alexandra},
  booktitle={Proceedings of the 54th annual meeting of the association for computational linguistics (volume 1: long papers)},
  pages={86--96},
  year={2016}
}

@article{scherer2004emotions,
  title={Emotions in everyday life: Probability of occurrence, risk factors, appraisal and reaction patterns},
  author={Scherer, Klaus R and Wranik, Tanja and Sangsue, Janique and Tran, V{\'e}ronique and Scherer, Ursula},
  journal={Social Science Information},
  volume={43},
  number={4},
  pages={499--570},
  year={2004},
  publisher={Sage Publications Sage CA: Thousand Oaks, CA}
}

@article{botev2017variance,
  title={Variance reduction},
  author={Botev, Zdravko and Ridder, Ad},
  journal={Wiley statsRef: Statistics reference online},
  pages={1--6},
  year={2017},
  publisher={John Wiley \& Sons, Ltd. Hoboken, NJ, USA}
}

@inproceedings{huggingface,
  title={Transformers: State-of-the-art natural language processing},
  author={Wolf, Thomas and Debut, Lysandre and Sanh, Victor and Chaumond, Julien and Delangue, Clement and Moi, Anthony and Cistac, Pierric and Rault, Tim and Louf, Remi and Funtowicz, Morgan and others},
  booktitle={Proceedings of the 2020 conference on empirical methods in natural language processing: system demonstrations},
  pages={38--45},
  year={2020}
}

@misc{ma2019nlpaug,
  title={NLP Augmentation},
  author={Edward Ma},
  howpublished={https://github.com/makcedward/nlpaug},
  year={2019}
}

@inproceedings{venigalla-emotion,
  title={Understanding emotions of developer community towards software documentation},
  author={Venigalla, Akhila Sri Manasa and Chimalakonda, Sridhar},
  booktitle={2021 IEEE/ACM 43rd International Conference on Software Engineering: Software Engineering in Society (ICSE-SEIS)},
  pages={87--91},
  year={2021},
  organization={IEEE}
}

@inproceedings{feng2021survey,
  title={A Survey of Data Augmentation Approaches for NLP},
  author={Feng, Steven Y and Gangal, Varun and Wei, Jason and Chandar, Sarath and Vosoughi, Soroush and Mitamura, Teruko and Hovy, Eduard},
  booktitle={Findings of the Association for Computational Linguistics: ACL-IJCNLP 2021},
  pages={968--988},
  year={2021}
}

@INPROCEEDINGS{eeshita-sentiment,
  author={Biswas, Eeshita and Karabulut, Mehmet Efruz and Pollock, Lori and Vijay-Shanker, K.},
  booktitle={2020 IEEE International Conference on Software Maintenance and Evolution (ICSME)}, 
  title={Achieving Reliable Sentiment Analysis in the Software Engineering Domain using BERT}, 
  year={2020},
  volume={},
  number={},
  pages={162-173}}

@article{li2022data,
  title={Data augmentation approaches in natural language processing: A survey},
  author={Li, Bohan and Hou, Yutai and Che, Wanxiang},
  journal={AI Open},
  year={2022},
  publisher={Elsevier}
}

@inproceedings{devlin2018bert,
  title={Bert: Pre-training of deep bidirectional transformers for language understanding},
  author={Devlin, Jacob and Chang, Ming-Wei and Lee, Kenton and Toutanova, Kristina},
  booktitle={Proceedings of the 2019 conference of the North American chapter of the association for computational linguistics: human language technologies, volume 1 (long and short papers)},
  pages={4171--4186},
  year={2019}
}

@inproceedings{lewis2019bart,
  title={BART: Denoising sequence-to-sequence pre-training for natural language generation, translation, and comprehension},
  author={Lewis, Mike and Liu, Yinhan and Goyal, Naman and Ghazvininejad, Marjan and Mohamed, Abdelrahman and Levy, Omer and Stoyanov, Veselin and Zettlemoyer, Luke},
  booktitle={Proceedings of the 58th annual meeting of the association for computational linguistics},
  pages={7871--7880},
  year={2020}
}

@inproceedings{wei-zou-2019-eda,
  title={EDA: Easy Data Augmentation Techniques for Boosting Performance on Text Classification Tasks},
  author={Wei, Jason and Zou, Kai},
  booktitle={Proceedings of the 2019 Conference on Empirical Methods in Natural Language Processing and the 9th International Joint Conference on Natural Language Processing (EMNLP-IJCNLP)},
  pages={6382--6388},
  year={2019}
}

@article{Guo2019AugmentingDW,
  title={Augmenting Data with Mixup for Sentence Classification: An Empirical Study},
  author={Hongyu Guo and Yongyi Mao and Richong Zhang},
  journal={ArXiv},
  year={2019},
  volume={abs/1905.08941}
}

@Manual{Syuzhet,
    title = {Syuzhet: Extract Sentiment and Plot Arcs from Text},
    author = {Matthew L. Jockers},
    year = {2015},
    url = {https://github.com/mjockers/syuzhet},
}

@article{lu2018first,
  title={A first look at emoji usage on github: An empirical study},
  author={Lu, Xuan and Cao, Yanbin and Chen, Zhenpeng and Liu, Xuanzhe},
  journal={arXiv preprint arXiv:1812.04863},
  year={2018}
}

@article{son2021more,
  title={More Than React: Investigating The Role of EmojiReaction in GitHub Pull Requests},
  author={Son, Teyon and Xiao, Tao and Wang, Dong and Kula, Raula Gaikovina and Ishio, Takashi and Matsumoto, Kenichi},
  journal={arXiv preprint arXiv:2108.08094},
  year={2021}
}

@inproceedings{kovatchev2021vectors,
  title={Can vectors read minds better than experts? Comparing data augmentation strategies for the automated scoring of children’s mindreading ability},
  author={Venelin Kovatchev and Phillip Smith and Mark G. Lee and Rory T. Devine},
  booktitle={ACL},
  year={2021}
}

@article{shorten2021text,
  title={Text data augmentation for deep learning},
  author={Shorten, Connor and Khoshgoftaar, Taghi M and Furht, Borko},
  journal={Journal of big Data},
  volume={8},
  number={1},
  pages={1--34},
  year={2021},
  publisher={Springer}
}

@INPROCEEDINGS{8595220,
  author={Novielli, Nicole and Girardi, Daniela and Lanubile, Filippo},
  booktitle={2018 IEEE/ACM 15th International Conference on Mining Software Repositories (MSR)}, 
  title={A Benchmark Study on Sentiment Analysis for Software Engineering Research}, 
  year={2018},
  pages={364-375}
}

@inproceedings{cbert,
  title={Conditional bert contextual augmentation},
  author={Wu, Xing and Lv, Shangwen and Zang, Liangjun and Han, Jizhong and Hu, Songlin},
  booktitle={International Conference on Computational Science},
  pages={84--95},
  year={2019},
  organization={Springer}
}

@article{xie2019unsupervised,
  title={Unsupervised data augmentation for consistency training},
  author={Xie, Qizhe and Dai, Zihang and Hovy, Eduard and Luong, Thang and Le, Quoc},
  journal={Advances in Neural Information Processing Systems},
  volume={33},
  pages={6256--6268},
  year={2020}
}

@inproceedings{novielli2018gold,
  title={A gold standard for emotion annotation in stack overflow},
  author={Novielli, Nicole and Calefato, Fabio and Lanubile, Filippo},
  booktitle={2018 IEEE/ACM 15th International Conference on Mining Software Repositories (MSR)},
  pages={14--17},
  year={2018},
  organization={IEEE}
}

@inproceedings{mixup,
  title={mixup: Beyond Empirical Risk Minimization},
  author={Zhang, Hongyi and Cisse, Moustapha and Dauphin, Yann N and Lopez-Paz, David},
  booktitle={International Conference on Learning Representations},
  year={2018}
}

@inproceedings{goemotions,
  title={GoEmotions: A Dataset of Fine-Grained Emotions},
  author={Demszky, Dorottya and Movshovitz-Attias, Dana and Ko, Jeongwoo and Cowen, Alan and Nemade, Gaurav and Ravi, Sujith},
  booktitle={Proceedings of the 58th Annual Meeting of the Association for Computational Linguistics},
  pages={4040--4054},
  year={2020}
}

@article {CowenE7900,
	author = {Cowen, Alan S. and Keltner, Dacher},
	title = {Self-report captures 27 distinct categories of emotion bridged by continuous gradients},
	volume = {114},
	number = {38},
	pages = {E7900--E7909},
	year = {2017}
}

@article{cowen2020face,
  title={What the face displays: Mapping 28 emotions conveyed by naturalistic expression.},
  author={Cowen, Alan S and Keltner, Dacher},
  journal={American Psychologist},
  year={2020},
  publisher={American Psychological Association}
}

@article{cowen2019mapping,
  title={Mapping 24 emotions conveyed by brief human vocalization.},
  author={Cowen, Alan S and Elfenbein, Hillary Anger and Laukka, Petri and Keltner, Dacher},
  journal={American Psychologist},
  volume={74},
  number={6},
  pages={698},
  year={2019},
  publisher={American Psychological Association}
}

@inproceedings{mrini-etal-2021-gradually,
  title={A gradually soft multi-task and data-augmented approach to medical question understanding},
  author={Mrini, Khalil and Dernoncourt, Franck and Yoon, Seunghyun and Bui, Trung and Chang, Walter and Farcas, Emilia and Nakashole, Ndapandula},
  booktitle={Proceedings of the 59th Annual Meeting of the Association for Computational Linguistics and the 11th International Joint Conference on Natural Language Processing (Volume 1: Long Papers)},
  pages={1505--1515},
  year={2021}
}

@article{opinionlitreview,
  title={Opinion mining for software development: a systematic literature review},
  author={Lin, Bin and Cassee, Nathan and Serebrenik, Alexander and Bavota, Gabriele and Novielli, Nicole and Lanza, Michele},
  journal={ACM Transactions on Software Engineering and Methodology (TOSEM)},
  volume={31},
  number={3},
  pages={1--41},
  year={2022},
  publisher={ACM New York, NY}
}

@article{Graziotin15,
author = {Graziotin, Daniel and Wang, Xiaofeng and Abrahamsson, Pekka},
title = {Do Feelings Matter? On the Correlation of Affects and the Self-Assessed Productivity in Software Engineering},
year = {2015},
issue_date = {July 2015},
publisher = {John Wiley &amp; Sons, Inc.},
address = {USA},
volume = {27},
number = {7},
issn = {2047-7473}
}

@inproceedings{Muller15,
  title={Stuck and frustrated or in flow and happy: Sensing developers' emotions and progress},
  author={M{\"u}ller, Sebastian C and Fritz, Thomas},
  booktitle={2015 IEEE/ACM 37th IEEE International Conference on Software Engineering},
  volume={1},
  pages={688--699},
  year={2015},
  organization={IEEE}
}

@inproceedings{Girardi20,
  title={Recognizing developers' emotions while programming},
  author={Girardi, Daniela and Novielli, Nicole and Fucci, Davide and Lanubile, Filippo},
  booktitle={Proceedings of the ACM/IEEE 42nd international conference on software engineering},
  pages={666--677},
  year={2020}
}

@inproceedings{Graziotin13,
  title={Are happy developers more productive? The correlation of affective states of software developers and their self-assessed productivity},
  author={Graziotin, Daniel and Wang, Xiaofeng and Abrahamsson, Pekka},
  booktitle={International Conference on Product Focused Software Process Improvement},
  pages={50--64},
  year={2013},
  organization={Springer}
}

@inproceedings{Murgia2014DoDF,
  title={Do developers feel emotions? an exploratory analysis of emotions in software artifacts},
  author={A. Murgia and Parastou Tourani and B. Adams and Marco Ortu},
  booktitle={MSR 2014},
  year={2014}
}

@article{Amabile2005AffectAC,
  title={Affect and Creativity at Work},
  author={T. Amabile and Sigal G. Barsade and J. S. Mueller and B. Staw},
  journal={Administrative Science Quarterly},
  year={2005},
  volume={50},
  pages={367 - 403}
}

@article{GRAZIOTIN201832,
  title={What happens when software developers are (un) happy},
  author={Graziotin, Daniel and Fagerholm, Fabian and Wang, Xiaofeng and Abrahamsson, Pekka},
  journal={Journal of Systems and Software},
  volume={140},
  pages={32--47},
  year={2018},
  publisher={Elsevier}
}

@article{8666786,
  title={Today was a good day: The daily life of software developers},
  author={Meyer, Andr{\'e} N and Barr, Earl T and Bird, Christian and Zimmermann, Thomas},
  journal={IEEE Transactions on Software Engineering},
  volume={47},
  number={5},
  pages={863--880},
  year={2019},
  publisher={IEEE}
}

@inproceedings{Graziotin17,
author = {Graziotin, Daniel and Fagerholm, Fabian and Wang, Xiaofeng and Abrahamsson, Pekka},
title = {On the Unhappiness of Software Developers},
year = {2017},
isbn = {9781450348041},
publisher = {Association for Computing Machinery},
address = {New York, NY, USA},
url = {https://doi.org/10.1145/3084226.3084242},
doi = {10.1145/3084226.3084242},
booktitle = {Proceedings of the 21st International Conference on Evaluation and Assessment in Software Engineering},
pages = {324–333},
numpages = {10},
location = {Karlskrona, Sweden},
series = {EASE'17}
}

@article{Forsgren21,
  title={The SPACE of Developer Productivity: There's more to it than you think.},
  author={Forsgren, Nicole and Storey, Margaret-Anne and Maddila, Chandra and Zimmermann, Thomas and Houck, Brian and Butler, Jenna},
  journal={Queue},
  volume={19},
  number={1},
  pages={20--48},
  year={2021},
  publisher={ACM New York, NY, USA}
}

@inproceedings{Imtiaz18,
author = {Imtiaz, Nasif and Middleton, Justin and Girouard, Peter and Murphy-Hill, Emerson},
title = {Sentiment and Politeness Analysis Tools on Developer Discussions Are Unreliable, but so Are People},
year = {2018},
isbn = {9781450357517},
publisher = {Association for Computing Machinery},
address = {New York, NY, USA}
}

@inproceedings{Sarker19,
author = {Sarker, Farhana and Vasilescu, Bogdan and Blincoe, Kelly and Filkov, Vladimir},
title = {Socio-Technical Work-Rate Increase Associates with Changes in Work Patterns in Online Projects},
year = {2019},
publisher = {IEEE Press}
}

@article{Kuutila2020ChatAI,
  title={Chat activity is a better predictor than chat sentiment on software developers productivity},
  author={Miikka Kuutila and M. M{\"a}ntyl{\"a} and Ma{\"e}lick Claes},
  journal={Proceedings of the IEEE/ACM 42nd International Conference on Software Engineering Workshops},
  year={2020}
}

@INPROCEEDINGS {Fucci21,
author = {G. Fucci and N. Cassee and F. Zampetti and N. Novielli and A. Serebrenik and M. Di Penta},
booktitle = {2021 2021 IEEE/ACM 18th International Conference on Mining Software Repositories (MSR) (MSR)},
title = {Waiting Around or job Half-Done? Sentiment in Self-Admitted Technical Debt},
year = {2021},
volume = {},
issn = {},
pages = {403-414},
doi = {10.1109/MSR52588.2021.00052},
url = {https://doi.ieeecomputersociety.org/10.1109/MSR52588.2021.00052},
publisher = {IEEE Computer Society},
address = {Los Alamitos, CA, USA},
month = {may}
}

@article{Ebert2021AnES,
  title={An exploratory study on confusion in code reviews},
  author={Felipe Ebert and Fernando Castor and Nicole Novielli and Alexander Serebrenik},
  journal={Empirical Software Engineering},
  year={2021},
  volume={26},
  pages={1-48}
}

@inproceedings{tabassum2020code,
    title = "Code and Named Entity Recognition in {S}tack{O}verflow",
    author = "Tabassum, Jeniya  and
      Maddela, Mounica  and
      Xu, Wei  and
      Ritter, Alan",
    booktitle = "Proceedings of the 58th Annual Meeting of the Association for Computational Linguistics",
    month = jul,
    year = "2020",
    address = "Online",
    publisher = "Association for Computational Linguistics"
}

@article{novielli2021assessment,
   title={Assessment of off-the-shelf SE-specific sentiment analysis tools: An extended replication study},
   volume={26},
   ISSN={1573-7616},
   url={http://dx.doi.org/10.1007/s10664-021-09960-w},
   DOI={10.1007/s10664-021-09960-w},
   number={4},
   journal={Empirical Software Engineering},
   publisher={Springer Science and Business Media LLC},
   author={Novielli, Nicole and Calefato, Fabio and Lanubile, Filippo and Serebrenik, Alexander},
   year={2021},
   month={Jun}
}

@article{tsoumakas2010random,
  title={Random k-labelsets for multilabel classification},
  author={Tsoumakas, Grigorios and Katakis, Ioannis and Vlahavas, Ioannis},
  journal={IEEE transactions on knowledge and data engineering},
  volume={23},
  number={7},
  pages={1079--1089},
  year={2010},
  publisher={IEEE}
}

@inproceedings{tsoumakas2008effective,
  title={Effective and efficient multilabel classification in domains with large number of labels},
  author={Tsoumakas, Grigorios and Katakis, Ioannis and Vlahavas, Ioannis},
  booktitle={Proc. ECML/PKDD 2008 Workshop on Mining Multidimensional Data (MMD’08)},
  volume={21},
  year={2008}
}

@article{cabrera2020classifying,
  title={Classifying emotions in Stack Overflow and JIRA using a multi-label approach},
  author={Cabrera-Diego, Luis Adri{\'a}n and Bessis, Nik and Korkontzelos, Ioannis},
  journal={Knowledge-Based Systems},
  volume={195},
  pages={105633},
  year={2020},
  publisher={Elsevier}
}

@inproceedings{emod,
  title={EmoD: An end-to-end approach for investigating emotion dynamics in software development},
  author={Neupane, Krishna Prasad and Cheung, Kabo and Wang, Yi},
  booktitle={2019 IEEE International Conference on Software Maintenance and Evolution (ICSME)},
  pages={252--256},
  year={2019},
  organization={IEEE}
}

@inproceedings{deepmoji,
  title={Using millions of emoji occurrences to learn any-domain representations for detecting sentiment, emotion and sarcasm},
  author={Felbo, Bjarke and Mislove, Alan and S{\o}gaard, Anders and Rahwan, Iyad and Lehmann, Sune},
  booktitle={Proceedings of the 2017 Conference on Empirical Methods in Natural Language Processing},
  pages={1615--1625},
  year={2017}
}

@article{chen2021emoji,
author = {Chen, Zhenpeng and Cao, Yanbin and Yao, Huihan and Lu, Xuan and Peng, Xin and Mei, Hong and Liu, Xuanzhe},
title = {Emoji-Powered Sentiment and Emotion Detection from Software Developers’ Communication Data},
year = {2021},
issue_date = {March 2021},
publisher = {Association for Computing Machinery},
address = {New York, NY, USA},
volume = {30},
number = {2},
issn = {1049-331X},
url = {https://doi.org/10.1145/3424308},
doi = {10.1145/3424308},
journal = {ACM Trans. Softw. Eng. Methodol.},
month = jan,
articleno = {18},
numpages = {48}
}

@inproceedings{Tourani,
author = {Tourani, Parastou and Jiang, Yujuan and Adams, Bram},
title = {Monitoring Sentiment in Open Source Mailing Lists: Exploratory Study on the Apache Ecosystem},
year = {2014},
publisher = {IBM Corp.},
address = {USA},
booktitle = {Proceedings of 24th Annual International Conference on Computer Science and Software Engineering},
pages = {34?44},
numpages = {11},
location = {Markham, Ontario, Canada},
series = {CASCON ?14}
}

@INPROCEEDINGS{Chatterjee21,
  author={Chatterjee, Preetha and Damevski, Kostadin and Pollock, Lori},
  booktitle={Proceedings of the 2021 IEEE/ACM 43rd International Conference on Software Engineering (ICSE)}, 
  title={Automatic Extraction of Opinion-based {Q\&A} from Online Developer Chats}, 
  year={2021},
  volume={},
  number={},
  pages={1260-1272},
  doi={10.1109/ICSE43902.2021.00115}}

@inproceedings{Chatterjee2020Journal,
 author = {P. Chatterjee and K. Damevski and N.A. Kraft and L. Pollock},
 title = {{Automatically Identifying the Quality of Developer Chats for Post Hoc Use}},
 booktitle = {Transactions on Software Engineering and Methodology (TOSEM)},
 series = {TOSEM '20},
 year = {2020},
 }

@inproceedings{Chatterjee2019,
  author    = {P. Chatterjee and K. Damevski and L. Pollock and V. Augustine and N.A. Kraft},
  title     = {{Exploratory Study of Slack Q\&A Chats as a Mining Source for Software Engineering Tools}},
  booktitle = {Proceedings of the 16th International Conference on Mining Software Repositories (MSR'19)},
  month     = may,
  year      = 2019,
 doi = {10.1109/MSR.2019.00075},
  location  = {Montreal, Canada},
}

@article{Senti4SD,
  title={Sentiment Polarity Detection for Software Development},
  author={Fabio Calefato and Filippo Lanubile and Federico Maiorano and Nicole Novielli},
  journal={Empirical Software Engineering},
  year={2017},
  volume={23},
  pages={1352-1382}
}

@inproceedings{ortu2016,
author = {Ortu, Marco and Murgia, Alessandro and Destefanis, Giuseppe and Tourani, Parastou and Tonelli, Roberto and Marchesi, Michele and Adams, Bram},
title = {The Emotional Side of Software Developers in JIRA},
year = {2016},
isbn = {9781450341868},
publisher = {Association for Computing Machinery},
address = {New York, NY, USA},
url = {https://doi.org/10.1145/2901739.2903505},
doi = {10.1145/2901739.2903505},
booktitle = {Proceedings of the 13th International Conference on Mining Software Repositories},
pages = {480–483},
numpages = {4},
location = {Austin, Texas},
series = {MSR '16}
}

@inproceedings{sanei2021impacts,
  title={The Impacts of Sentiments and Tones in Community-Generated Issue Discussions},
  author={Sanei, Arghavan and Cheng, Jinghui and Adams, Bram},
  booktitle={2021 IEEE/ACM 13th International Workshop on Cooperative and Human Aspects of Software Engineering (CHASE)},
  pages={1--10},
  year={2021},
  organization={IEEE}
}

@ARTICLE{8802324,
  author={Novielli, Nicole and Serebrenik, Alexander},
  journal={IEEE Software}, 
  title={Sentiment and Emotion in Software Engineering}, 
  year={2019},
  volume={36},
  number={5},
  pages={6-23}}

@incollection{ekman,
	author = {Paul Ekman},
	year = {1999},
	title = {Basic Emotions},
	pages = {4--5},
	booktitle = {Handbook of Cognition and Emotion},
	editor = {Tim Dalgleish and M. J. Powers},
	publisher = {Wiley}
}

@article{bujang2017simplified,
  title={A Simplified Guide to Determination of Sample Size Requirements for Estimating the Value of Intraclass Correlation Coefficient: a Review.},
  author={Bujang, Mohamad Adam and Baharum, Nurakmal},
  journal={Archives of Orofacial Science},
  volume={12},
  number={1},
  year={2017}
}

@article{stemler2004comparison,
  title={A comparison of consensus, consistency, and measurement approaches to estimating interrater reliability},
  year={2004},
  author={Stemler, Steven E}
}

@inproceedings{zhang2021supporting,
  title={Supporting Clustering with Contrastive Learning},
  author={Zhang, Dejiao and Nan, Feng and Wei, Xiaokai and Li, Shang-Wen and Zhu, Henghui and Mckeown, Kathleen and Nallapati, Ramesh and Arnold, Andrew O and Xiang, Bing},
  booktitle={Proceedings of the 2021 Conference of the North American Chapter of the Association for Computational Linguistics: Human Language Technologies},
  pages={5419--5430},
  year={2021}
}

@article{radford2019language,
  title={Language models are unsupervised multitask learners},
  author={Radford, Alec and Wu, Jeffrey and Child, Rewon and Luan, David and Amodei, Dario and Sutskever, Ilya and others},
  journal={OpenAI blog},
  volume={1},
  number={8},
  pages={9},
  year={2019}
}

@ARTICLE{gan-model-overview,
  author={Creswell, Antonia and White, Tom and Dumoulin, Vincent and Arulkumaran, Kai and Sengupta, Biswa and Bharath, Anil A.},
  journal={IEEE Signal Processing Magazine}, 
  title={Generative Adversarial Networks: An Overview}, 
  year={2018},
  volume={35},
  number={1},
  pages={53-65},
  doi={10.1109/MSP.2017.2765202}}

@inproceedings{baccianella2010sentiwordnet,
  title={Sentiwordnet 3.0: An enhanced lexical resource for sentiment analysis and opinion mining},
  author={Baccianella, Stefano and Esuli, Andrea and Sebastiani, Fabrizio},
  booktitle={Proceedings of the Seventh International Conference on Language Resources and Evaluation (LREC'10)},
  year={2010}
}

@article{russell1977evidence,
  title={Evidence for a three-factor theory of emotions},
  author={Russell, James A and Mehrabian, Albert},
  journal={Journal of research in Personality},
  volume={11},
  number={3},
  pages={273--294},
  year={1977},
  publisher={Elsevier}
}

@article{zhang2015character,
  title={Character-level convolutional networks for text classification},
  author={Zhang, Xiang and Zhao, Junbo and LeCun, Yann},
  journal={Advances in neural information processing systems},
  volume={28},
  year={2015}
}

@article{miller1995wordnet,
  title={WordNet: a lexical database for English},
  author={Miller, George A},
  journal={Communications of the ACM},
  volume={38},
  number={11},
  pages={39--41},
  year={1995},
  publisher={ACM New York, NY, USA}
}

\end{document}